\documentclass[preprint,authoryear,12pt]{elsarticle}
\usepackage{comment}
\usepackage{float}
\usepackage{booktabs}
\usepackage[table]{xcolor}
\usepackage{multirow}
\usepackage{booktabs} 
\usepackage{subcaption}
\usepackage{url}
\usepackage[skins]{tcolorbox}
\usepackage{hyperref}
\tcbset{
    sharp corners,
    colback = white,
    before skip = 0.2cm,    
    after skip = 0.5cm      
}   
\definecolor{main}{HTML}{5989cf}    
\definecolor{sub}{HTML}{cde4ff}     
\newtcolorbox{boxK}{
    sharpish corners, 
    boxrule = 0pt,
    toprule = 2.5pt, 
    enhanced,
    fuzzy shadow = {0pt}{-2pt}{-0.5pt}{0.5pt}{black!35} 
}

\usepackage{amssymb}
\usepackage{amsmath}

\journal{Nuclear Physics B}

\begin{document}

\begin{frontmatter}



\title{PRIME-SVR: Physics-infoRmed Implicit Multi-Echo Slice-to-Volume Reconstruction for Fetal T2 mapping}


\author[1,2]{Busra Bulut\corref{cor1}}
\ead{busra.bulut@unil.ch}

\author[3]{Maik Dannecker}

\author[1,2]{Thomas Sanchez}

\author[6]{Sara Neves Silva}

\author[7]{Steven Jia}

\author[1,2]{Jean-Baptiste Ledoux}
\author[10]{Leo Pomar}
\author[10]{Joanna Sichitiu}
\author[10]{Yvan Gomez}
\author[1]{Meriam Koob}
\author[1]{Vincent Dunet}
\author[6]{Maria Deprez}

\author[7]{Guillaume Auzias}

\author[8]{François Rousseau}

\author[4,9]{Jana Hutter}

\author[3,4,5]{Daniel Rueckert}

\author[1,2]{Meritxell Bach Cuadra}

\affiliation[1]{organization={Department of Medical Radiology, Lausanne University Hospital and University of Lausanne},
            city={Lausanne},
            country={Switzerland}}

\affiliation[2]{organization={CIBM Center for Biomedical Imaging},
            city={Lausanne},
            country={Switzerland}}

\affiliation[3]{organization={Chair for AI in Healthcare and Medicine, Technical University of Munich (TUM) and TUM University Hospital},
            city={Munich},
            country={Germany}}

\affiliation[4]{organization={Department of Computing, Imperial College London},
            city={London},
            country={United Kingdom}}

\affiliation[5]{organization={Munich Center for Machine Learning (MCML)},
            city={Munich},
            country={Germany}}

\affiliation[6]{organization={Biomedical Engineering Department, School of Biomedical Engineering and Imaging Sciences, King's College London},
            city={London},
            country={United Kingdom}}

\affiliation[7]{organization={Institut de Neurosciences de la Timone, UMR 7289, CNRS, Aix-Marseille Université},
            city={Marseille},
            postcode={13005},
            country={France}}

\affiliation[8]{organization={IMT Atlantique, LaTIM UMR 1101},
            city={Brest},
            country={France}}

\affiliation[9]{organization={Leibniz Universität Hannover},country={Germany}}

\affiliation[10]{organization={Department Woman-Mother-Child, Lausanne University Hospital and University of Lausanne},
            city={Lausanne},
            country={Switzerland}}

\cortext[cor1]{Corresponding author}

\begin{abstract}
Slice-to-volume reconstruction (SVR) has advanced fetal brain MRI by reconstructing high-resolution (HR) volumes from motion-corrupted thick 2D stacks of slices. However, reconstructing beyond the narrow clinical echo time (TE) range remains unsolved. Established SVR methods are validated and optimized only within this narrow range, which restricts voxel-wise quantitative T2 mapping, a promising biomarker for fetal brain assessment that requires acquisitions spanning a wider range of TEs. Here we introduce PRIME-SVR, the first implicit neural representation framework for joint multi-TE SVR. PRIME-SVR is fully self-supervised: one network maps spatial coordinates to HR intensities across TEs, while a second estimates slice motion. Cross-TE coherence is enforced through a Bloch equation-derived regularization that constrains the reconstruction to the mono-exponential T2 decay subspace, without requiring ground-truth T2 values. An adaptive weighting scheme further strengthens this coupling for degraded stacks while leaving high-quality data reconstruction-driven. We demonstrate PRIME-SVR on the largest multi-echo fetal MRI dataset to date, 117 stacks from 39 in vivo acquisitions (13 subjects × 3 TEs) across two centers, two vendors, and two field strengths (1.5 T and 0.55 T). PRIME-SVR outperforms established SVR baselines, improving reconstruction sharpness (AES) by 47\%, anatomical accuracy (SSIM) by 30\%, and cross-TE structural consistency (NMI, Edge Dice) by 14\%, relative to the best-performing baseline per metric and field strength. We further demonstrate its clinical value by producing the first 0.8 mm isotropic T2 maps at 0.55 T and the first T2 maps derived from INR-based SVR. Finally, we show that PRIME-SVR yields high-quality T2 maps even when acquisition time is reduced from 15 to 5 minutes, paving the way for deploying self-supervised T2 mapping directly on the scanner.
\end{abstract}

\end{frontmatter}



\section{Introduction}
\label{sec1}

Fetal magnetic resonance imaging (MRI) has become an essential complementary tool to ultrasound because of its excellent soft-tissue contrast \citep{wataganara_fetal_2016}. 
In particular, T2 weighted (T2w) imaging is suited to visualize microstructural changes in both white matter (WM) and grey matter (GM) reflecting normal fetal brain maturation (neuronal migration, myelination) or pathological conditions such as congenital Cytomegalovirus (CMV) infection. T2w imaging is consequently a valuable tool for assessing fetal brain development \textit{in utero}, thus having a significant impact on pregnancy management \citep{dubois_early_2014, yarnykh_quantitative_2018, griffiths_use_2017}. However, fetal MRI faces major acquisition challenges linked to unpredictable fetal motion and maternal breathing. To mitigate these artifacts, current clinical acquisition strategies rely on Single-Shot T2w (SST2w) sequences \citep{prayer_fetal_2004}: stacks of thick 2D slices (typically 3–4.5 mm at 1.5~T or 0.55~T) are acquired in an interleaved manner, fast enough to effectively freeze intra-slice motion, although inter-slice motion still occurs. Acquisitions are conventionally obtained in at least 3 orthogonal orientations (sagittal, coronal, and axial) of the fetal brain (see Fig.\ref{fig:princ} A). In a next step, these stacks can then be reconstructed into an isotropic high resolution (HR) 3D volume using slice-to-volume reconstruction (SVR) \citep{rousseau_registration-based_2006,gholipour_robust_2010,kuklisova-murgasova_reconstruction_2012, tourbier_efficient_2015}. Usually, more than 3 stacks (averaging 6 to 12) are acquired to improve reconstruction quality, always requiring careful balance with scan duration for maternal comfort and cost effectiveness. However, like standard weighted clinical sequences, these reconstructed volumes provide relative contrast for qualitative visualization rather than absolute measurements. Consequently, reconstructed images remain highly dependent on acquisition protocols and centers, and this is further compounded by the choice of SVR algorithm, which introduces a domain shift between reconstructed volumes, making the harmonization of multi-centric fetal brain MRI data particularly challenging \citep{sanchez_biometry_2025}.

Transitioning to quantitative MRI, specifically T2 mapping, helps mitigate this limitation via a voxel-wise estimation of the T2 relaxation time, i.e., the time constant governing signal decay across varying TEs \citep{bloch_nuclear_1946} (Fig. \ref{fig:princ} D). While some residual dependence on center specifics and protocols persists, T2 mapping significantly reduces these biases and provides a more objective measure of tissue properties than qualitative visualization at a single TE, crucial for characterizing both typical and atypical fetal brain development \citep{bhattacharya_vivo_2024}.
It can also support the standardization of perinatal MRI, linking prenatal and postnatal imaging. In fetal MRI, due to practical constraints (motion and scan duration), conventional multi-echo spin-echo (MESE) acquisition sequences \citep{whittall_vivo_1997}, primarily selected for adult and neonatal T2 mapping, cannot be employed. Prior work has instead achieved T2 mapping by repeating the SST2w sequence at several TEs, with independent SVR reconstruction performed at each TE \citep{lajous_t2_2020, bhattacharya_vivo_2024, roulet_t2_2025}.

However, this strategy is suboptimal for several reasons. First, existing SVR methods have been developed and validated on contrasts corresponding to clinical TEs while the most challenging reconstructions are outside of this range, where the signal-to-noise ratio (SNR) is the lowest \citep{joshi_optimization_2026}. Second, reconstructing each TE independently leads to a loss of coherence in reconstructed volumes and a suboptimal use of available information of the same subject \citep{bulut_physics-informed_2025}. Third, independent processing can alter the intensity distribution unevenly across TEs, disrupting the signal decay curve and rendering downstream T2 fitting unreliable. These limitations motivate our work for a unified SVR framework specifically designed to handle multi-TE acquisitions jointly. By leveraging the complementary information across TEs, our method improves reconstruction quality, enables T2 mapping on challenging data and reduces the number of stacks required per TE when acquisition quality is sufficiently high, shortening total scan time from approximately 15 to 5 minutes.

As T2 relaxometry is field-strength dependent, a robust validation of the tool should take such a dimension into account.
Beyond the 1.5~T field strength commonly used in fetal exams, we are particularly interested in lower field strength.
Indeed, mid-field MRI (0.55~T) exhibits longer T2 decay, enhancing sensitivity to short-T2 tissues such as deep gray matter (DGM) \citep{payette_t2_2025, marques_lowfield_2019} and reducing T2-induced blurring in SST2w imaging \citep{qin_point_2012}. Reduced specific absorption rate (SAR) constraints at mid-field further allow the unconstrained use of 180° refocusing pulses \citep{krishnamurthy_mr_2015}, better isolating true spin-spin relaxation so improving the accuracy of T2 quantification. Mid-field systems also offer practical advantages in cost and deployability, supporting quantitative fetal imaging in resource-limited settings \citep{arnold_lowfield_2023} and a wider bore for pregnant women. However, the reduced SNR often necessitates thicker slice acquisitions, increasing the ill-posedness of SVR, and the shift in image contrast can degrade post-processing tools trained on higher-field data, such as brain mask extraction \citep{faghihpirayesh2024fetal} or slice pre-alignment \citep{xu_svort_2022}. These challenges motivate the need for an SVR algorithm for T2 mapping to be robust to field strength variations. To date, in-vivo fetal brain T2 mapping at 0.55~T remains largely unexplored, with only a preliminary proof-of-concept on two subjects reported in our prior workshop work \citep{bulut_physics-informed_2025}.

\subsection{Contributions}

This work extends our preliminary workshop study \citep{bulut_physics-informed_2025}, which demonstrated the feasibility of the framework on adult acquisitions with simulated fetal-like motion and on simulated fetal brain data. The present work introduces automated adaptive regularization and validates the extended framework on acquired multi-echo fetal MRI data. The main contributions are as follows:

\begin{itemize}

\item We introduce PRIME-SVR, a fully self-supervised, physics-informed implicit neural representation (INR) framework for slice-to-volume reconstruction that jointly reconstructs multi-echo fetal brain volumes acquired at three TEs within a unified model, thereby enabling quantitative T2 relaxometry.

\item We propose an automated adaptive weighting strategy for the physics-based regularization derived from the Bloch equations. The regularization strength is determined from stack-level quality metrics computed directly from the input data, such that stronger cross-TE coupling is applied to more severely degraded stacks.

\item We validate PRIME-SVR on newly acquired multi-echo fetal MRI data collected at two centers and two field strengths, 0.55~T and 1.5~T, together with the five fetal cases previously reported in the literature. This study considerably expands the number of available fetal brain T2 maps, presents the first such maps at 0.55~T, and provides the first within-subject comparison of fetal brain T2 maps acquired at both 0.55~T and 1.5~T.

\end{itemize}

\newpage

\section{Related works}

\subsection{Classical SVR methods at a single contrast}
\label{sec:SVR}
SVR is the inverse problem of recovering a 3D volume from motion-corrupted thick 2D slices acquired at different orientations. In the fetal brain, the forward model, also referred as the slice acquisition model, accounts for slice-wise rigid motion and the point spread function (PSF), representing blurring and downsampling; solving the SVR problem consists of estimating slice motion parameters (slice-to-volume registration) and reconstructing the HR volume (super-resolution reconstruction), commonly through an iterative scheme alternating between the two. Early methods differed mainly in their similarity metrics and PSF approximation \citep{rousseau_registration-based_2006, jiang_mri_2007}. A key development was the introduction of outlier-robust reconstruction by \citet{gholipour_robust_2010}, who used an M-estimator-based error norm to reduce the influence of corrupted slices, later extended by \citet{kuklisova-murgasova_reconstruction_2012} into a statistically driven framework (SVRTK) capable of fully discarding corrupted slices. Subsequent work improved computational efficiency via multi-GPU implementations \citep{kainz_fast_2015} and reconstruction quality via total variation regularization \citep{tourbier_efficient_2015}, while \citet{ebner_automated_2020} proposed a fully automated pipeline integrating brain localization, segmentation, and slice-level outlier rejection. Despite these advances, recent studies \citep{uus_retrospective_2022, ciceri_geometric_2023, sanchez_biometry_2025, joshi_optimization_2026} have shown that SVR effectiveness remains dependent on the degree of fetal motion, the number of available stacks, and the resolution of the HR volume.

\subsubsection{Slice motion estimation}
To address the challenges posed by moderate to severe fetal motion, several recent works have focused on estimating slice motion prior to SVR reconstruction. In particular, learning-based methods that directly infer rigid transformations from acquired slices have emerged as powerful initialization strategies for modern SVR frameworks. \citet{uus_scanner-based_2025} proposed a deep-learning-based pose initialization approach in which a 3D U-Net predicts fetal brain landmarks from low-resolution MRI stacks. These landmarks are used to estimate a rigid transformation into a standard anatomical space, providing an initialization for the SVRTK reconstruction pipeline. In parallel, the transformer-based SVoRT framework \citep{xu_svort_2022} models inter-slice spatial relationships to directly estimate rigid slice transformations. Despite their strong performance, this can degrade on unseen data, like any supervised machine learning technique. This is particularly limiting for multi-TE acquisitions, where TE beyond the clinical ranges can be used.

\subsubsection{Implicit Neural Representation for Fetal SVR at a single contrast}
\label{sec:INR}
SVR can also be formulated as a volume rendering problem, enabling the use of emerging implicit neural representation (INR) techniques. The HR volume is represented as a continuous function of spatial coordinates describing the underlying brain anatomy. This function is approximated by a multi-layer perceptron (MLP) and learned in a self-supervised manner for each subject using the acquired stacks of slices. As a result, the reconstruction is not constrained to a predefined voxel grid and can be queried at arbitrary spatial locations, making it resolution-agnostic \citep{wu_irem_2021}. \citet{xu_nesvor_2023} adapt this method for fetal data by proposing NeSVoR. They combined the MLP with a hash-grid encoding for multi scale feature extraction \citep{muller_instant_2022} and used the low-frequency components to correct bias fields. In addition, voxel variance of slices are used as an outlier coefficients during reconstruction. Slice motion parameters are treated as learnable variables but it is usually initialized using an external motion estimation algorithm such as SVoRT \citep{xu_svort_2022}. More recently, \citet{dannecker_meta-learning_2025} introduced a framework based on two sinusoidal representation networks (SIRENs), i.e., MLPs with sinusoidal activation functions \citep{sitzmann_implicit_2020}, for fetal SVR. The first network represents the continuous 3D anatomy, similarly to NeSVoR, whereas the second models a continuous mapping from slice coordinates to slice-specific parameters, including slice motion and outlier coefficients. This formulation yields a fully self-supervised reconstruction pipeline that does not rely on an external slice motion estimation. 
A recent alternative implicit representation is Gaussian splatting \citep{kerbl_3d_2023}, in which the volume is modeled as a finite set of spatially distributed Gaussian primitives rather than a neural coordinate function. \citet{dannecker2026fast} adapted this approach for fetal SVR and demonstrated reconstruction quality comparable to NeSVoR while substantially reducing training time. 
Although INR-based methods have substantially improved fetal brain SVR, they have been developed and evaluated primarily for qualitative reconstruction limited clinical range of TEs. Their applicability to quantitative MRI remains unexplored.

\subsection{Multi-contrast SVR methods}

By learning a continuous representation of the underlying anatomy, INRs are naturally well suited for multi-contrast reconstruction. \citet{mcginnis_single-subject_2023} proposed the first INR-based multi-contrast SVR framework for adult brain imaging, combining Fourier feature encoding \citep{tancik_fourier_2020} with a split-head MLP to jointly model multiple contrasts, though without slice motion estimation or outlier rejection. \citet{link-sourani_joint_2025} later adapted this framework to fetal SVR. In both studies, contrasts were limited to T1w and T2w images, with cross-contrast relationships learned only implicitly by the network rather than explicitly modeled. Extension to multi-TE contrasts remains largely unaddressed: \citet{beirinckx_model-based_2022} proposed a classical (non-INR) SVR framework estimating HR 3D T2 maps directly from MESE stacks in adults, but its iterative scheme is highly sensitive to T2 and motion initialization, limiting it to settings with minimal inter-slice motion and artifacts, unsuitable for fetal imaging. More recently, \citet{bhattacharya2025joint} explored joint registration of multi-TE stacks prior to SVRTK reconstruction in a preliminary abstract, confirming community interest in this direction.

\subsection{Fetal Brain T2 Relaxometry with SST2w}
 \citet{lajous_t2_2020} and \citet{roulet_t2_2025} validated SST2w sequence's accuracy on T2 relaxometry at 1.5~T and 0.55~T using phantom and adult in vivo data (without SVR), both reporting a systematic T2 overestimation attributable to the sequence itself. To our knowledge, \citet{bhattacharya_vivo_2024} is the only study to report T2 maps from in vivo SST2w acquisition on 5 fetuses after SVR reconstruction (at 1.5~T). They use SVRTK with the second-TE stack as registration template and mitigating slice-profile-induced overestimation via extended phase graph (EPG) \citep{weigel_extended_2015} based dictionary fitting; however, this strategy does not guarantee successful reconstruction at late TEs nor anatomical continuity across TEs, as shown in our preliminary work \citep{bulut_physics-informed_2025}. That workshop paper was the first to derive T2 maps from INR-based SVR and at 0.55~T, quantifying using adult in vivo data with simulated fetal-like motion and signal dropouts that multi-TE SVR does not bias T2 estimates relative to single-TE SVR while improving reconstruction quality. 

\newpage
\section{Method}
\begin{figure}[h]
    \centering
    \makebox[\textwidth][c]{%
        \includegraphics[width=1\textwidth]{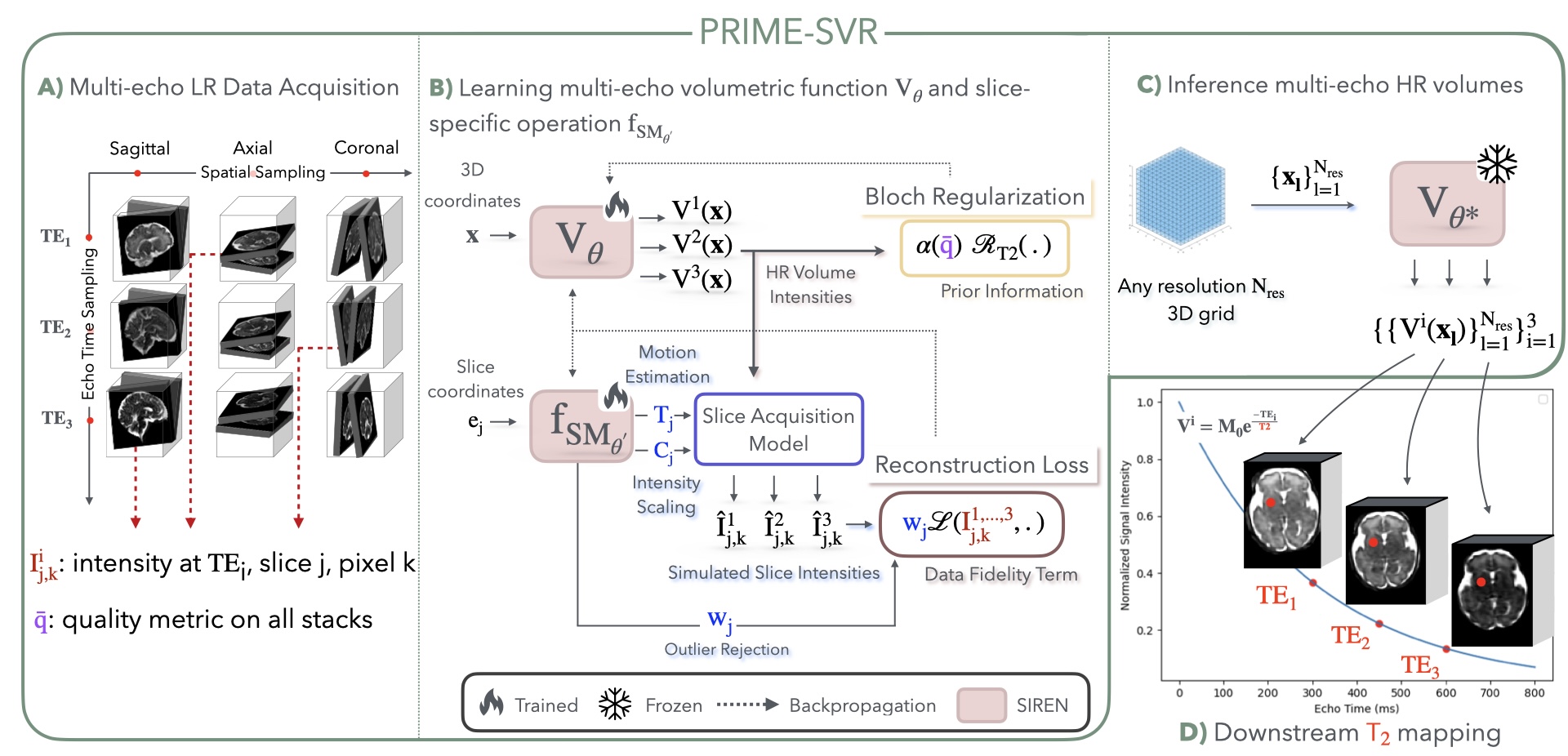}
     }
    \caption{\textbf{Overview of the proposed method PRIME-SVR}. Multi-echo Low Resolution stacks are acquired across orthogonal orientations and TEs; a quality metric $\bar{q}$ on all stacks is estimated. (B) A shared SIREN $\mathbf{V_\theta}$ maps spatial coordinates to HR intensities at each TE, while $\mathbf{f_{SM_\theta}}$ predicts slice-specific acquisition parameters. Training combines a data-fidelity term with a Bloch regularization, adaptively weighted by $\alpha(\bar{q})$. (C) Once trained, $\mathbf{V_\theta}$ is queried on an arbitrary-resolution grid to produce HR multi-TE volumes. (D) HR volumes are fit to the exponential decay model to obtain T2 maps.}
    \label{fig:princ}
\end{figure}

In PRIME-SVR, we model the multi-echo HR intensities as an implicit continuous function $\mathbf{V}$ defined over 3D spatial coordinates. The multi-echo data observed in the acquired 2D slices are treated as sparse, discrete, and degraded samples of this underlying function. The degradation process is slice-specific and modeled through a slice acquisition model, where a subset of its parameters is estimated by a separate implicit continuous function $\mathbf{f_{SM}}$ defined over the coordinates within the slice. Both functions are parameterized by $\theta$ and $\theta'$, respectively, and are learned in a self-supervised manner for each subject by minimizing the discrepancy between the observed slices and those simulated through the slice acquisition model. In addition, a regularization term derived from the Bloch equations and weighted by $\alpha(\bar{q})$ is imposed on $\mathbf{V}_{\theta}$ to model the physical relationship across TEs. Once learned, $\mathbf{V}{_\theta}$ can be queried on any grid sampling the 3D domain to reconstruct high-resolution multi-echo volumes, from which a quantitative T2 map is estimated (see Fig.~\ref{fig:princ}).

\subsection{Sinusoidal Decomposition of $\mathbf{V_{\theta}}$  }

We approximate the unknown continuous function $\mathbf{V}$ by a parametrized function $\mathbf{V_\theta}$ defined as
\[
\mathbf{V_\theta} :
\begin{aligned}
\mathbb{R}^3 &\to \mathbb{R}^{N_{\mathrm{TE}}} \\
\mathbf{x} &\mapsto \mathbf{V_\theta(\mathbf{x})} =
\left[
V^1_\theta(\mathbf{x}),
\dots,
V^{N_{\mathrm{TE}}}_\theta(\mathbf{x})
\right],
\end{aligned}
\]
where \(V^i_\theta(\mathbf{x})\) denotes the HR volume intensity at $\mathrm{TE}_i$ and spatial coordinate \(\mathbf{x}\), and $N_{\mathrm{TE}}$ is the number of TEs. We model $\mathbf{V}_{\theta}$ using a sinusoidal representation network (SIREN)~\citep{sitzmann_implicit_2020}, i.e., a multilayer perceptron (MLP) with sinusoidal activation functions. The network outputs \(N_{\mathrm{TE}}\) channels, corresponding to the set of TEs. We assume that this architecture is sufficiently expressive to approximate $\mathbf{V}$ with high accuracy. More precisely, \(\mathbf{V_\theta}\) is defined as a composition of sinusoidal layers:
\[
\mathbf{V_\theta(\mathbf{x})} =
W_L \, \sin\!\Big(
W_{L-1}\sin\!\big(
\dots
\sin\!\big(
W_1\mathbf{x} + b_1
\big)
\dots
\big) + b_{L-1}
\Big) + b_L,
\]
where \(\theta = \{W_\ell, b_\ell\}_{\ell=1}^{L}\) are learnable parameters, and the sine activation is applied element-wise. This representation is well-suited for modeling high-frequency spatial variations while preserving smoothness \citep{essakine_where_2024}. Moreover, it does not require explicit positional encoding, as the sinusoidal activations provide a rich representation of the input space. It is worth noting that all channels share the same hidden layers and relationships across TEs are therefore learned implicitly through the shared network representation and its learnable parameters.

\subsection{Multi-Echo Continuous Slice Acquisition Model}
\label{subsection:model}

To relate the continuous HR multi-echo representation
$\mathbf{V}$ to the acquired low-resolution slices, we define a continuous slice acquisition model. Let
$\mathbf{I} \in \mathbb{R}^{N_{\mathrm{s}} \times N_{\mathrm{p}}}$ denote the acquired set of thick slices, where
$N_{\mathrm{s}}$ is the total number of acquired slices over all stacks and all TEs, and
$N_{\mathrm{p}}$ is the number of pixels per slice.
The slices are organized into stacks, each corresponding to a particular orientation with high in-plane resolution; we do not explicitly distinguish between stacks here. We denote by
\[
\tau(j) \in \{1,\dots,N_{\mathrm{TE}}\}
\]
the echo index associated with slice $j$, i.e.\ the TE at which slice $j$ was acquired, with $N_{\mathrm{TE}}$ the number of TEs defined previously. Thus $I_{j,k}$ denotes the intensity of pixel $k$ in slice $j$, implicitly acquired at $\mathrm{TE}_{\tau(j)}$. Following \citet{xu_nesvor_2023}, the continuous slice acquisition model is:
\begin{align}
I_{j,k}
&=
C_j
\int_{\Omega}
M_{j,k}(\mathbf{x}) \,
B_j(\mathbf{x}) \,
\left[
V^{\tau(j)}(\mathbf{x}) + \varepsilon_j(\mathbf{x})
\right]
\, \mathrm{d}\mathbf{x} .
\label{eq:continuous_forward}
\end{align}
Combining with the mono-exponential T2 relaxation model derived from the Bloch equations~\citep{brown_magnetic_2014},

\[
V^{\tau(j)}(\mathbf{x})
=
M_0(\mathbf{x})
\exp\!\left(
-\mathrm{TE}_{\tau(j)} / \mathrm{T2}(\mathbf{x})
\right),
\]
the slice acquisition model can be written as:
\begin{align}
I_{j,k}
&=
C_j
\int_{\Omega}
M_{j,k}(\mathbf{x}) \,
B_j(\mathbf{x}) \,
\left[
M_0(\mathbf{x}) e^{-\mathrm{TE}_{\tau(j)}/\mathrm{T2}(\mathbf{x})}
+ \varepsilon_j(\mathbf{x})
\right]
\, \mathrm{d}\mathbf{x},
\label{eq:continuous_forward_echo}
\end{align}
where $\Omega \subset \mathbb{R}^3$ denotes the continuous 3D region of interest, $C_j$ is an unknown scaling factor accounting for global intensity inconsistencies of slice $j$, $M_{j,k}(\cdot)$ is the anisotropic Gaussian point spread function (PSF) centered at $p_{j,k}$, the spatial location corresponding to pixel $k$ in slice $j$ (see \ref{sec:annexe1}), $B_j(\cdot)$ is the unknown bias field at slice $j$, $V^{\tau(j)}(\cdot)$ is the unknown HR volume at $\mathrm{TE}_{\tau(j)}$, $M_0(\cdot)$ is the initial transverse magnetization and $\mathrm{T2}(\cdot)$ is the T2 relaxometry map, and $\varepsilon_j(\cdot)$ denotes the residual noise term, assumed to follow white Gaussian noise, with $\mathbb{E}[\varepsilon_j(x)] = 0\text{ and }\mathbb{E}[\varepsilon_j(x)\varepsilon_j(y)]={\sigma_j}^2\delta(x - y),$ where $\delta(\cdot)$ is the Dirac delta function.

\subsection{Estimation of Slice-Specific Parameters: $\mathbf{f_{SM_{\theta}}}$}
To regress the slice-specific parameters appearing in Eq.~(\ref{eq:continuous_forward}) from the slice coordinate, we deploy a second SIREN network. The bias field $B_j(\cdot)$ is handled as part of the preprocessing. We define
\[ \mathbf{f_{{SM}_{\theta}}} : \begin{aligned} [-1,1] &\to \mathbb{R}^{6} \times \mathbb{R} \times \mathbb{R} \\ e_j &\mapsto [T_j, C_j, w_j] \end{aligned}, \]
where $e_j$ is the normalized slice index over all slices across all stacks and TEs. $T_j$ denotes the rigid transformation parameters (rotation and translation) mapping slice $j$ into the HR 3D space, needed to define $M_{j,k}$ (see \ref{sec:annexe1}), $C_j$ is defined previously, and $w_j$ is an outlier coefficient. 
In contrast to~\cite{xu_nesvor_2023}, where this coefficient is linked to the variance estimation, here $w_j$ is treated as an independent learnable parameter. To prevent the trivial solution $w_j \to 0$, we introduce a logarithmic barrier regularization term \citep{boyd2004convex} that penalizes vanishing weights. In addition, using $\tau(\cdot)$ to identify slices sharing the same TE, we constrain the normalization of these parameters within each TE: for every $i \in \{1,\dots,N_{\mathrm{TE}}\}$, letting $\mathcal{S}_i = \{j : \tau(j) = i\}$ and $N_s^i = |\mathcal{S}_i|$,

\[
\frac{1}{N_s^i}\sum_{j \in \mathcal{S}_i} w_j = 1, \qquad \frac{1}{N_s^i}\sum_{j \in \mathcal{S}_i} C_j = 1.
\]

\subsection{Training strategy}

Although the multi-echo continuous slice acquisition model provides a comprehensive description of the image formation process, the mono exponential model remains an useful but simple approximation that may not capture the full complexity of the acquired signals. Directly embedding this simplified physical model into the forward operator could therefore propagate model mismatch to the reconstructed volumes and introduce systematic bias. We instead decouple the physical signal model from the acquisition process and incorporate it as prior information by introducing
\begin{itemize}
    \item a data fidelity term enforcing consistency with the slice acquisition model defined in Eq.~(\ref{eq:continuous_forward});
    \item a regularization term enforcing temporal consistency across TEs according to Bloch equation, weighted by stack-specific acquisition quality metrics.
\end{itemize}

\paragraph{Data Fidelity}

We focus on solving Eq.~\eqref{eq:continuous_forward}. The expectation of the pixel intensity can therefore be written as
\begin{equation} \mathbb{E}(I_{j,k}) = C_j \int_{\Omega} M_{j,k}(\mathbf{x}) \, V^{\tau(j)}(\mathbf{x}) \, \mathrm{d}\mathbf{x}. \end{equation}
This integral does not admit a closed-form solution. Following~\citet{xu_nesvor_2023}, we approximate it using Monte Carlo integration. 
Let $\mathbf{u}_\ell \sim \mathcal{N}(\mathbf{0},\boldsymbol{\Sigma})$, for $\ell=1,\dots,L$ with $\boldsymbol{\Sigma}$ defined in \ref{sec:annexe1}. The forward model becomes
\begin{equation}
\hat{I}_{j,k}
\approx
C_j
\frac{1}{L}
\sum_{\ell=1}^{L}
V^{\tau(j)}\!\big(
T_j(\mathbf{p}_{j,k}+\mathbf{u}_\ell)
\big).
\label{eq:expectedI}
\end{equation}
\paragraph{Bloch Regularization}
We incorporate the mono-exponential T2 relaxation model derived from the Bloch equations as prior knowledge by enforcing that the reconstructed multi-echo HR volume intensities lie within the subspace spanned by exponential functions after each iteration during training. To achieve this, we minimize the residual between the logarithm of the reconstructed volumes intensities and their projection onto this exponential subspace. Taking the logarithm of the mono exponential model yields a linear relationship :
\[
\log V^{i}(\mathbf{x})
=
\log M_0(\mathbf{x})
-
\frac{\mathrm{TE}_i}{\mathrm{T2}(\mathbf{x})}.
\]
Stacking all TEs defines
\[
y(\mathbf{x}) =
\begin{bmatrix}
\log V^{1}(\mathbf{x}) \\
\vdots \\
\log V^{N_{\mathrm{TE}}}(\mathbf{x})
\end{bmatrix},
\quad
D =
\begin{bmatrix}
1 & \mathrm{TE}_1 \\
\vdots & \vdots \\
1 & \mathrm{TE}_{N_{\mathrm{TE}}}
\end{bmatrix},
\quad
\beta(\mathbf{x}) =
\begin{bmatrix}
\log M_0(\mathbf{x}) \\
-1/\mathrm{T2}(\mathbf{x})
\end{bmatrix}.
\]
This yields the linear model
\[
y(\mathbf{x}) = D\,\beta(\mathbf{x}),
\]
which implies that the log-signal vector $y(\mathbf{x})$ lies in the two-dimensional subspace spanned by the columns of $D$. For distinct TEs, $D$ has full column rank and the orthogonal projection onto this subspace is
\[
P = D(D^\top D)^{-1}D^\top.\]
We quantify the deviation from the Bloch signal model as the squared norm of the component orthogonal to this subspace:
\[
r_{\mathrm{T2}}(y(\mathbf{x})) =
\| (P-I)\,y(\mathbf{x}) \|_2^2.
\]
The Bloch regularization computed on a set $X$ of coordinates becomes
\begin{equation}
R_{\mathrm{T2}}(V^1,\dots,V^{N_{\mathrm{TE}}})
=
\frac{1}{|X|}
\sum_{\mathbf{x}\in X}
\|(P-I)\,y(\mathbf{x})\|_2^2.
\label{eq:reg}
\end{equation}
Since $P-I$ depends only on the TEs, it can be precomputed and the voxel-wise evaluation of Eq.~\eqref{eq:reg} does not introduce additional computational complexity compared to the data fidelity term.

\subsubsection{Loss Function}
$\mathbf{f_{{SM}_{\theta}'}}$ and \(\mathbf{V_\theta}\)  are jointly optimized over batches $\mathcal{B}$ of pixel $k$ in slice $j$ via gradient descent. The Bloch regularization is computed on the set $X$, where each element is the spatial coordinate
$\mathbf{x}_{j,k} = T_j(\mathbf{p}_{j,k})$ of a pixel in the batch. We use the mean absolute error to reduce the impact of outliers. The overall loss function is defined as
\begin{equation}
\label{eq:loss_function_sc}
\begin{split}
\mathcal{L}(V^1,\dots,V^{N_{\mathrm{TE}}})
= &
\frac{1}{|\mathcal{B}|}
\sum_{(j,k)\in\mathcal{B}} w_j
\left| I_{j,k} - \hat{I}_{j,k} \right| \\ &+
\alpha(\bar{q})\, R_{\mathrm{T2}}(V^1,\dots,V^{N_{\mathrm{TE}}})
+ \gamma \frac{1}{|\mathcal{B}|} \sum_{j\in\mathcal{B}} -\log(w_j).
\end{split}
\end{equation}

Here, $\gamma \in \mathbb{R}, \quad 0 < \gamma < 1$ is a regularization parameter controlling the logarithmic barrier term of the outlier rejection $w$ defined previously. $\alpha(\cdot)$ is a function balancing the Bloch regularization term defined in the following paragraph.

\paragraph{Quality-adaptive weight} Since the Bloch regularization can introduce bias when the acquisition deviates from the idealized signal model, its influence should increase only when the observed stacks contain strong artifacts. In such cases, stronger coupling across TEs helps compensate for missing or corrupted information. We therefore adapt $\alpha$ based on the predicted stacks quality score $q$ provided by FetMRQC~\citep{sanchez_fetmrqc_2024}. It extracts an ensemble of quality metrics from fetal SST2w acquisitions and combines them to predict experts’ ratings using random forests. Let $\bar{q}$ denote the mean quality score across all stacks of all TEs.

\[
\alpha(\bar{q})=
\begin{cases}
[10,30], & \text{if } \bar{q}<0.9 \text{ or less than 3 stacks/TE},\\
]0,1], & \text{if } \bar{q}\ge 0.9.
\end{cases}
\]
This quality-adaptive schedule increases the contribution of the physical prior for low-quality acquisitions while preserving data-driven reconstruction for high-quality stacks. The effect of $\alpha$ on the T2 distribution and this design choice are analyzed in the ablation study.

\subsection{Inference and T2 Fitting Strategies}
\label{sec:t2}
Once trained, $\mathbf{V_\theta}$ can be queried on a 3D grid, producing multi-echo HR volumes. As detailed in Section~\ref{subsection:model}, the SST2w acquisition scheme requires a T2 fitting that explicitly accounts for noise in the signal model. We use the same closed-form approximation as in the Bloch regularization to initialize the T2 and $M_0$. This estimation is refined using the L-BFGS-B algorithm \citep{byrd_limited_1995}, a quasi-Newton method with bound constraints, following the same parameter settings as in~\cite{roulet_t2_2025}: T2 is bounded within $[10, 2000]$~ms, $M_0$ is bounded within $[\mathrm{signal}(\mathrm{TE}_{1}), 10000]$, and solver tolerances are set to $10^{-2}$ with a maximum of 50 line-search steps.

\section{Experimental setup}
\subsection{Datasets}
\begin{figure}[H]
    \centering
    \makebox[\textwidth][c]{%
        \includegraphics[width=\textwidth]{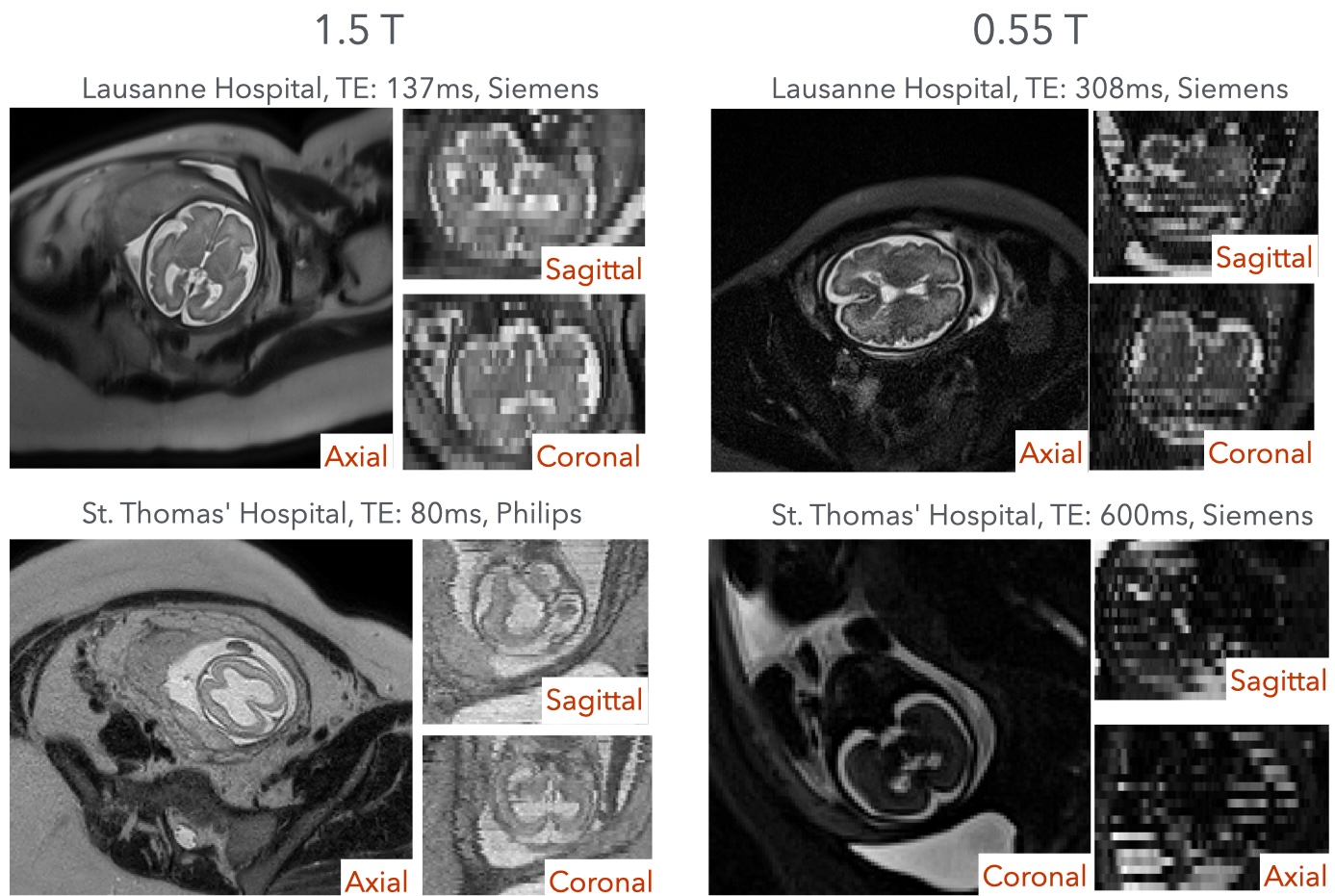}
    }
    \caption{\textbf{Illustrative examples of the multi-site, multi-field-strength dataset.} Representative low-resolution stacks (axial, sagittal, coronal orientations) acquired at 1.5~T (left) and 0.55~T (right) across two clinical sites (Lausanne Hospital and St.\ Thomas' Hospital), spanning different scanner vendors (Siemens, Philips) and a wide range of TEs.}
    \label{fig:OverviewData}
\end{figure}

Our framework is evaluated on the largest multi-echo fetal MRI dataset reported to date, comprising 117 stacks from 39 in-vivo fetal acquisitions (13 subjects $\times$ 3 TEs). The data were acquired at two centers from two vendors and spanning two field strengths (0.55~T and 1.5~T). Detailed acquisition parameters are provided in Table~\ref{tab:datasets}, and representative samples from each dataset are shown in Figure~\ref{fig:OverviewData}.

\begin{itemize}
    \item \textbf{In-vivo \textit{mid-field} fetal brain data (Site 1):} 
    Data were prospectively acquired for this study from five fetuses (GA = 24, 32, 32, 33, 37 weeks) using SST2w sequences at St. Thomas' Hospital, London, on a Siemens FreeMax 0.55~T scanner. The study was conducted under the ethically approved MEERKAT [REC: 21/LO/0742] and NANO [REC: 22/YH/0210] projects with data sharing explicitly included in the prospective ethics.

    \item \textbf{In-vivo \textit{high-field} fetal brain data (Site 1):} 
    Data from \cite{bhattacharya_vivo_2024} with five fetuses (GA = 21, 27, 29, 31, 35 weeks) acquired using SST2w sequences at St. Thomas' Hospital, London, on a Philips 1.5~T scanner. 

    \item  \textbf{In-vivo fetal brain data (Site 2):}
Data were prospectively acquired for this study from 2 fetuses at Lausanne University Hospital using SST2w sequences on Siemens FreeMax and Sola. One fetus was scanned at 0.55~T only, while the other was scanned at both 0.55~T and 1.5~T, with a 10-day interval between the two field-strength acquisitions. This study was approved by the ethics committee of the Canton of Vaud, Switzerland (project number: 2024-04995).

\end{itemize}

The choice of TE differed across acquisition settings. At 1.5~T, it was selected close to \cite{bhattacharya_vivo_2024}, while accounting for SAR constraints \citep{krishnamurthy_mr_2015}. At 0.55~T, TEs were selected to maximize sensitivity to the expected longer range of T2 (approximately 300--400 ms). Indeed, under a mono-exponential decay model, the Fisher information for estimating T2 is maximized when TE = T2 (see \ref{sec:annexe2}). Note, the clinical TE is around 110 ms \citep{ponrartana_low-field_2023}.


\begin{table}[h]
\centering
\caption{Summary of dataset acquisition parameters.}
\label{tab:datasets}

\renewcommand{\arraystretch}{1.2}
\setlength{\tabcolsep}{3pt}
\resizebox{\linewidth}{!}{
\begin{tabular}{cccccc}
\toprule
\textbf{\shortstack{Dataset\\Field}}  & \textbf{Vendor} & \textbf{\shortstack{Resolution \\ ($mm^3$)}} & \textbf{\shortstack{Slice Gap \\ (mm)}} & \textbf{\shortstack{TEs \\(ms)}} & \textbf{\shortstack{Participants \\ (GA in weeks)}} \\
\midrule

\multicolumn{6}{l}{\textbf{Site 1 : London}} \\
\midrule
0.55~T & Siemens & 1.48 $\times$ 1.48 $\times$ 4.5 & 4.5 & 300, 397, 600 & 5 (24, 32, 32, 33, 37) \\
1.5~T  & Philips &  1.21 $\times$ 1.21 $\times$ 2.5 & 1.25 & 80, 180, 400 & 5 (21, 27, 29, 31, 35) \\

\midrule
\multicolumn{6}{l}{\textbf{Site 2 : Lausanne}} \\
\midrule
0.55~T & Siemens &   1.18 $\times$ 1.18 $\times$ 4.5 & 4.5 & 308, 396, 598 & 2 (32, 33) \\
1.5~T  & Siemens &  1.18 $\times$ 1.18 $\times$ 4.5 & 4.5 & 137, 202, 400 & 1 (33) \\

\bottomrule
\end{tabular}}
\end{table}

\paragraph{Pre-processing}
All stacks were denoised using the non-local means algorithm implemented in Advanced Normalization Tools (ANTs) \citep{avants2009ants}, based on the method introduced by \cite{manjon_adaptive_2010}, followed by N4 bias field correction \citep{tustison_n4itk_2010}, also implemented in ANTs. Fetal brain masking was then performed using FET-BET \citep{faghihpirayesh2024fetal}; in cases where automated masking failed, manual corrections were applied using ITK-SNAP \citep{yushkevich_itk-snap_2016}.

\subsection{Implementation Details}
\label{sec:implementationdetails}
For the SIREN network modeling $\mathbf{V_{\theta}}$, we used six hidden layers with 330 units per layer. The initialization scheme proposed by \cite{sitzmann_implicit_2020} was adopted to stabilize training with sinusoidal activations. 
For the SIREN network modeling $\mathbf{f_{SM_{\theta}}}$, we used three hidden layers with 128 units per layer. Input coordinates were normalized, and a batch size of 25,000 was used and an epoch size of 100. Optimization was performed using the Adam optimizer with learning rates of $5 \times 10^{-4}$ for $\mathbf{f_{SM_{\theta}}}$ and $5 \times 10^{-5}$ for $\mathbf{V_{\theta}}$. $\gamma$ is set to 0.1 empirically. Training was conducted on a 48GB NVIDIA RTX 6000 GPU. The source code is available on GitHub\footnote{\url{https://github.com/Medical-Image-Analysis-Laboratory/PRIME-SVR}}.

\subsection{Comparative SVR methods}
We compare PRIME-SVR against 2 baseline methods :
\begin{itemize}
\item NeSVoR\footnote{\url{https://hub.docker.com/r/junshenxu/nesvor/tags}} \citep{xu_nesvor_2023}: an INR-based fetal SVR framework that jointly optimizes the reconstructed volume and slice-motion parameters. We used NeSVoR's automatic initialization option, which selects between SVoRT-based slice-motion estimation \citep{xu_svort_2022} and stack-to-stack registration. This option was used because SVoRT may generalize poorly to images acquired at TEs outside the range represented in its training data. Each TE is independently reconstructed. We also varied the hash-grid encoding resolution (coarser and finer multi-resolution levels than the default) in an attempt to adapt NeSVoR to the lower SNR of 0.55~T data; this did not yield an improvement across subjects. NeSVoR was therefore run with its default parameters, on the same workstation and GPU described in Section~\ref{sec:implementationdetails}.


\item SVRTK\footnote{\url{https://hub.docker.com/r/fetalsvrtk/svrtk/tags}} \citep{kuklisova-murgasova_reconstruction_2012}: a widely used CPU-based SVR framework based on iterative optimization. To optimize SVRTK for our low-field data, we first tested auto-SVRTK \citep{uus_scanner-based_2025}, a variant using CNN-based landmark detection trained on 0.55~T data; however, it failed to reconstruct our data, likely due to unmet input requirements (5--6 stacks per TE, TE range 80--180~ms), so we reverted to the standard pipeline. Multi-TE acquisitions were reconstructed following the protocol of \citet{bhattacharya_vivo_2024} (i.e. using the second-TE stack as the common registration template and without intensity matching) with 32 threads on an Intel Core i9-13900K CPU.

\end{itemize}

\subsection{Post-processing and T2 fitting}

All the reconstructions were done at 0.8 mm isotropic resolution.
For PRIME-SVR and SVRTK, the cross-TE HR volumes are already co-registered with one another as part of the reconstruction process. Rigid registration (B-spline interpolation, ANTs \citep{avants2009ants}) was therefore performed to align these volumes into the atlas space \citep{gholipour_normative_2017}. For NeSVoR, HR volumes are already reconstructed directly in the atlas space, but not aligned across TEs; the same rigid registration was applied to bring the multi-TE volumes into voxel-wise correspondence. This step is critical, as T2 fitting is performed voxel-wise and therefore requires highly accurate alignment across TEs. T2 fitting was performed using the algorithm described in Section~\ref{sec:t2}. Segmentation of the HR volumes was computed with FetalSynthSeg \citep{zalevskyi2026evaluating}.

\newpage

\section{Experiments}

We conducted 3 experiments and 2 ablation studies to evaluate PRIME-SVR on our multi-echo fetal dataset  : 

\subsection{Experiment 1: Reconstruction quality}
\begin{boxK}
\noindent\textbf{Hypothesis:} Joint multi-TE reconstruction improves HR image quality compared to baseline SVR, particularly with strong motion, at late TE and at 0.55~T.
\end{boxK}

To test this hypothesis, we compared PRIME-SVR with the baseline methods in terms of HR reconstruction quality at each TE. Since no ground truth was available, the evaluation included: (i) quantitative comparison between the acquired slices and slices resampled from the HR volumes capturing both structural similarity and the accuracy of motion estimation. All methods were evaluated using a common slice outlier mask to ensure a fair comparison, (ii) measurement of reconstructed volume sharpness, and (iii) structural consistency of HR reconstructions across TEs within the same subject.

\subsection{Experiment 2: T2 mapping}
\begin{boxK}
\noindent\textbf{Hypothesis:} PRIME-SVR's reconstruction yields more robust T2 estimates than baseline SVR, and enables reliable T2 mapping in subjects for whom independent per-TE reconstruction fails.

\end{boxK}

To test this hypothesis, we compare T2 maps derived from HR reconstructions of baseline SVR and PRIME-SVR.
The comparison covers: (i) the accuracy of the voxel-wise fit, measured by the residual fitting error, (ii) the resulting T2 values and their intra-region variability in white matter (WM) and deep gray matter (DGM).

\subsection{Experiment 3: Robustness to limited input data}
\begin{boxK}
\noindent\textbf{Hypothesis:} PRIME-SVR can obtain HR reconstruction and accurate T2 estimation when multi-echo input data are incomplete.
\end{boxK}

We simulated reduced-data acquisition scenarios by discarding stacks from the full data set (3 orthogonal stacks per TE). Specifically, we tested two reduced configurations: (i) 2 stacks/TE, where only 2 of the 3 orthogonal orientations is kept for each TE, and (ii) 1 stack/TE, where 1 of the 3 orthogonal orientations is kept for each TE. We aimed to evaluate whether the missing planes for a given TE can be compensated for by using stacks acquired at another TE, without compromising T2 accuracy. To assess this, we reconstructed HR images from both reduced and full data (denoted HR-reduced and HR-full, respectively) and evaluated (i) the structural similarity between HR-reduced and HR-full, and (ii) the mean absolute error in the T2 maps derived from HR-reduced versus HR-full, computed separately for WM and DGM.

\subsection{Ablation analysis} We performed a series of ablation studies to quantify the contribution of the main components of PRIME-SVR. 

\paragraph{Multi-TE coupling ablation}
We analyzed the respective contributions of the two mechanisms used to couple information across TEs: the shared multi-echo INR (implemented as a multi-channel network) and the Bloch regularization. We compared the reconstruction sharpness and structural consistency across TEs obtained with three configurations: (a) Reg-SVR, independent networks per TE with Bloch regularization coupling, (b) Shared-SVR, a shared multi-echo network without Bloch regularization, and (c) PRIME-SVR, the full model combining a shared network with Bloch regularization.

\paragraph{Adaptive weighting for Bloch regularization}
We investigated the importance of adapting the Bloch regularization weight rather than using a fixed value across all subjects. To this end, we compared the reconstruction sharpness,  structural consistency across TEs, and the mean and standard deviation of the estimated T$_2$ values in WM and DGM for reconstructions obtained using several fixed regularization weights and the proposed adaptive weighting strategy.

\subsection{Evaluation Metrics}

\paragraph{Structural Similarity (SSIM)}
SSIM, introduced by \cite{wang2004image}, evaluates perceptual image quality by jointly considering luminance, contrast, and structural similarity. It is defined as:
\begin{equation*}
\mathrm{SSIM}(x,y) = \frac{(2\mu_x \mu_y + C_1)(2\sigma_{xy} + C_2)}{(\mu_x^2 + \mu_y^2 + C_1)(\sigma_x^2 + \sigma_y^2 + C_2)},
\end{equation*}
where $\mu_x$ and $\mu_y$ denote the mean intensities, $\sigma_x^2$ and $\sigma_y^2$ the variances, and $\sigma_{xy}$ the covariance between images $x$ and $y$. Constants $C_1$ and $C_2$ are used to stabilize the division. 
    
\paragraph{Normalized Mutual Information (NMI)} is used to quantify the statistical dependence between two variables while remaining robust to intensity variations. We used the formulation proposed by \citet{studholme_1999_overlap}, defined as:
\[
\mathrm{NMI}(X,Y) = \frac{H(X) + H(Y)}{H(X,Y)},
\]
where $H(X)$, $H(Y)$, and $H(X,Y)$ denote the marginal and joint entropies, respectively. This formulation yields values in the range $[1,2]$, where $1$ indicates statistical independence and $2$ corresponds to perfect dependence.


\paragraph{Average Edge Strength (AES)}It is a gradient-based, reference-free metric used to quantify image blurring at anatomical edges \citep{pannetier_quantitative_2016}. Lower AES values indicate increased blurring, as edge gradients become less pronounced. \cite{eichhorn_evaluating_nodate} shows a strong correlation (Spearman's correlation of 0.6) of this metric with radiologists' visual assessment of image quality in turbo spin-echo acquisitions, like the SST2w sequence considered in this study. 
For slice $k$ in orientation $o$, AES is defined as:
\begin{align*}
AES^{(k,o)} =
\frac{
\sqrt{
\sum_{i,j} E\left(I_{ij}^{(k,o)}\right)
\left[
\left(G_x\left(I_{ij}^{(k,o)}\right)\right)^2 +
\left(G_y\left(I_{ij}^{(k,o)}\right)\right)^2
\right]
}
}{
\sum_{i,j} E\left(I_{ij}^{(k,o)}\right)
},
\end{align*}
where $I_{ij}^{(k,o)}$ denotes the image intensity at pixel location $(i,j)$ in slice $k$ of orientation $o$, $G_x$ and $G_y$ are centered gradient kernels along x and y, and $E\left(I_{ij}^{(k,o)}\right)$ is a binary edge mask obtained using the Canny edge detector \citep{canny_computational_1986}. AES for the volume is then computed as the average of $AES^{(k,o)}$ over all slices $k$ and all three orientations $o$.

\paragraph{Dice Score}
The Dice similarity coefficient (DSC) \citep{dice1945measures} is a widely used metric for evaluating spatial overlap between two binary masks $A$ and $B$. It is defined as:
\begin{equation*}
\mathrm{DSC} = \frac{2 |A \cap B|}{|A| + |B|}.
\end{equation*}
In this work, in addition to NMI, we assess structural consistency across TEs using the Dice overlap of Canny edge maps (Edge Dice) within each slice, applying $A$ and $B$ to the binary edge masks obtained at two different TEs. Unlike anatomical segmentation, whose performance is domain-dependent and may vary within TE, Canny edge detection is a deterministic method applied consistently across TEs, providing a robust measure of boundary-level structural agreement.

\subsection{Statistical Analysis}
\label{sec:stat}
Given the paired, repeated-measures design, each subject was reconstructed with every method under comparison, significance was assessed with the exact two-sided Wilcoxon signed-rank test on subject-level averages for reconstruction quality (averaged over TEs for AES, and over TE-pairs for NMI and Edge Dice) and subject-level region means (WM, DGM) for T2 accuracy (T2 values, residual fit), computed separately for each field strength. This non-parametric test was chosen over a paired t-test because of the small sample sizes (n=6 at 1.5 T, n = 7 at 0.55 T, n = 13 pooled for the ablation).

\newpage

\section{Results}

\subsection{Experiment 1: Reconstruction quality}
\label{exp:exp1}
Table~\ref{tab:srr_combined} summarizes the reconstruction quality and cross-TE structural consistency achieved by NeSVoR, SVRTK, and PRIME-SVR. Across both field strengths, PRIME-SVR achieved the highest slice-to-volume consistency (SSIM), with the largest improvement observed at 0.55~T, where SSIM improved from 0.55 to 0.82 over the best-performing baseline; this gain was statistically significant at both field strengths (Sec.~\ref{sec:stat}). Likewise, PRIME-SVR consistently produced the sharpest reconstructions, significantly increasing AES from 2.25 to 3.53 at 0.55~T and 2.47 to 4.09 at 1.5~T relative to the best baseline. For cross-TE consistency, PRIME-SVR obtained the highest NMI across all TE pairs, with the improvement reaching significance at both field strengths. Edge Dice revealed larger differences between methods than NMI. At 0.55~T, PRIME-SVR significantly improved cross-TE edge overlap from 0.43 to 0.63 across all TE pairs; at 1.5~T, PRIME-SVR's performance was comparable to NeSVoR, and this difference was not statistically significant, consistent with the smaller gap between methods at the higher field strength.

\begin{table}[bp]
\centering
\caption{Reconstruction quality and cross-TE structural consistency across SVR methods. \textbf{Top:} SSIM, mean $\pm$ std averaged across TEs. \textbf{Middle:} Average Edge Strength (AES, $\times 10^{-3}$), a reference-free sharpness metric, mean $\pm$ std averaged across TEs. \textbf{Bottom:} Normalized Mutual Information (NMI) and Canny edge dice across TE pairs, mean $\pm$ std, measuring cross-TE consistency. Best performance per row in \textbf{bold}.}
\label{tab:srr_combined}
\resizebox{\linewidth}{!}{
\begin{tabular}{llccc}
\toprule
Field Str.& Metric & \textbf{NeSVoR} & \textbf{SVRTK} & \textbf{PRIME-SVR (Ours)} \\
\midrule
\multirow{4}{*}{0.55~T} & SSIM $\uparrow$      & $0.47 \pm 0.25$ & $0.55 \pm 0.32$ & $\mathbf{0.82 \pm 0.11}^{*\dagger}$ \\
 & AES $\uparrow$       & $1.93 \pm 0.53$ & $2.25 \pm 1.32$ & $\mathbf{3.53 \pm 0.83}^{*\dagger}$ \\
                            & NMI $\uparrow$       & $1.30 \pm 0.03$ & $1.27 \pm 0.12$ & $\mathbf{1.37 \pm 0.04}^{*\dagger}$ \\
                            & Edge Dice $\uparrow$ & $0.43 \pm 0.04$ & $0.29 \pm 0.17$ & $\mathbf{0.63 \pm 0.10}^{*\dagger}$ \\
\midrule
\multirow{4}{*}{1.5~T}& SSIM $\uparrow$      & $0.57 \pm 0.25$ & $0.70 \pm 0.14$ & $\mathbf{0.77 \pm 0.11}^{*}$ \\
 & AES $\uparrow$       & $2.96 \pm 0.92$ & $2.47 \pm 0.89$ & $\mathbf{4.09 \pm 1.30}^{*\dagger}$ \\
                            & NMI $\uparrow$       & $1.36 \pm 0.02$ & $1.36 \pm 0.03$ & $\mathbf{1.41 \pm 0.04}^{*}$ \\
                            & Edge Dice $\uparrow$ & $0.60 \pm 0.07$ & $0.43 \pm 0.11$ & $0.60 \pm 0.10^{\dagger}$ \\
\bottomrule
\multicolumn{2}{l}{Reconstruction Time**} & 21 min& 2 min & 5 min \\
\bottomrule
\end{tabular}
}
\footnotesize{$^{*}$/$^{\dagger}$ $p \leq 0.05$, exact two-sided Wilcoxon signed-rank test, paired by subject, comparing PRIME-SVR against NeSVoR ($^{*}$) and SVRTK ($^{\dagger}$) separately (see Sec.~5.6); $n=7$ at 0.55~T, $n=6$ at 1.5~T.} 
\footnotesize{**Total reconstruction time for 3 TEs HR reconstruction. NeSVoR and PRIME-SVR were run on an NVIDIA RTX 6000 Ada Generation GPU with 48 GB of memory, SVRTK was run using 32 threads on an Intel Core i9-13900K CPU.}
\end{table}

\begin{figure}[htbp]
    \centering
    \vspace{-2cm}
    \includegraphics[width=0.7\textwidth]{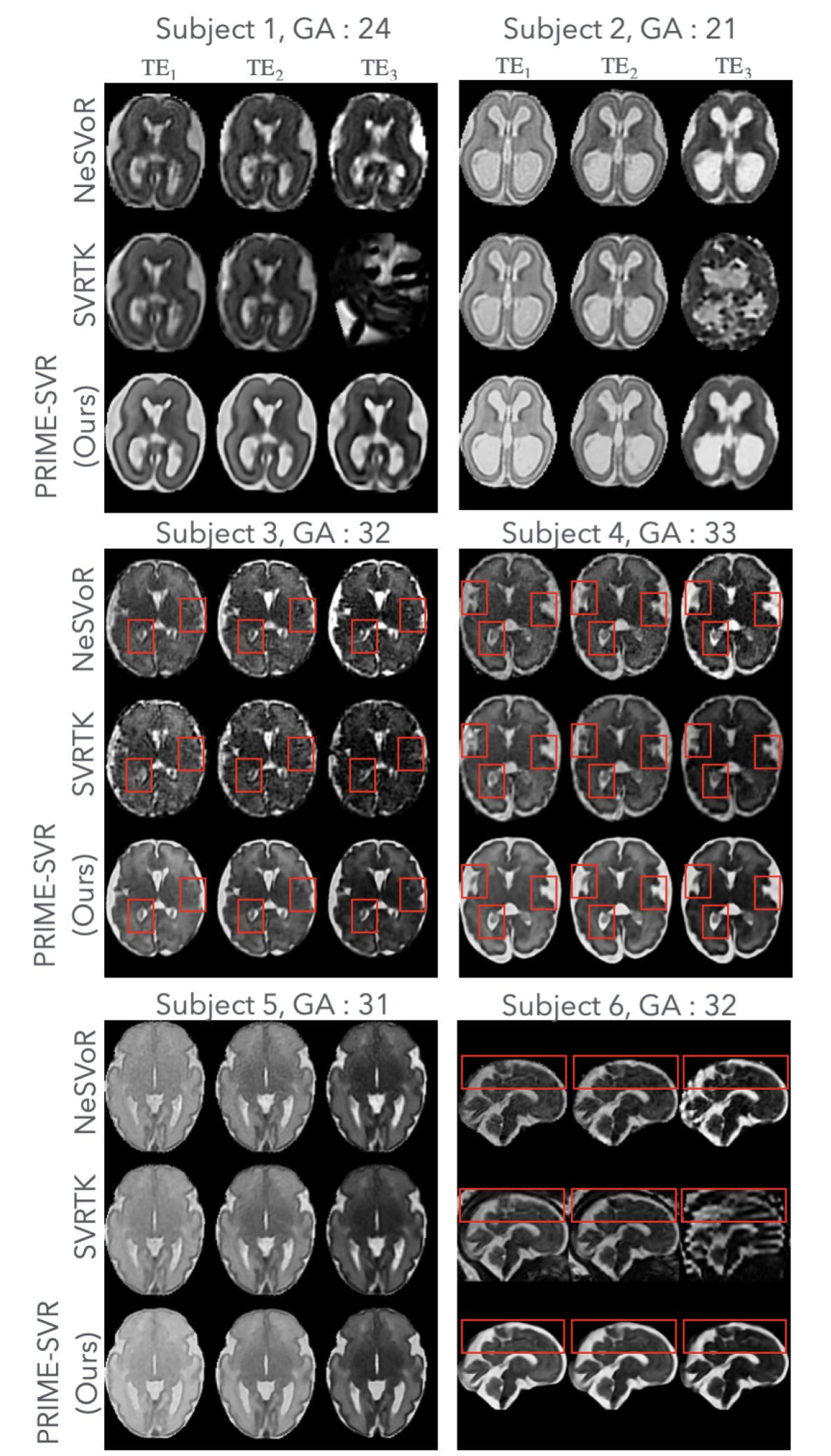}
    \caption{Overview for 6 subjects, the HR reconstructions (0.8mm isotropic resolution) at the 3 TEs for different SVR methods. Subject 2 and 5 were acquired at 1.5~T, while the rest of subjects were acquired at 0.55~T.}
    \vspace{-2cm}
    \label{fig:OverviewSRR}
\end{figure}

Figure~\ref{fig:OverviewSRR} illustrates reconstruction quality across six representative subjects, spanning both the most favorable and the most challenging cases in the dataset. At 0.55~T (Subjects 1, 3, 4, and 6), inter-method differences are most apparent: SVRTK fails to accurately reconstruct the latest TE acquisitions (Subjects 1 and 2), producing pronounced artifacts, consistent with its lower slice-to-volume consistency reported in Table 2. Even when all methods reconstruct the volume (Subjects 3, 4 and 6), PRIME-SVR preserves cortical boundaries and fine anatomical structures more consistently across TEs, as highlighted by the red boxes. Subject 5, acquired at 1.5~T, represents a favorable case where all three methods produce visually comparable reconstructions, in line with the smaller quantitative differences observed at this field strength.

\begin{figure}[bhpt]
    \vspace{-1cm}
    \centering
    \includegraphics[width=0.7\textwidth]{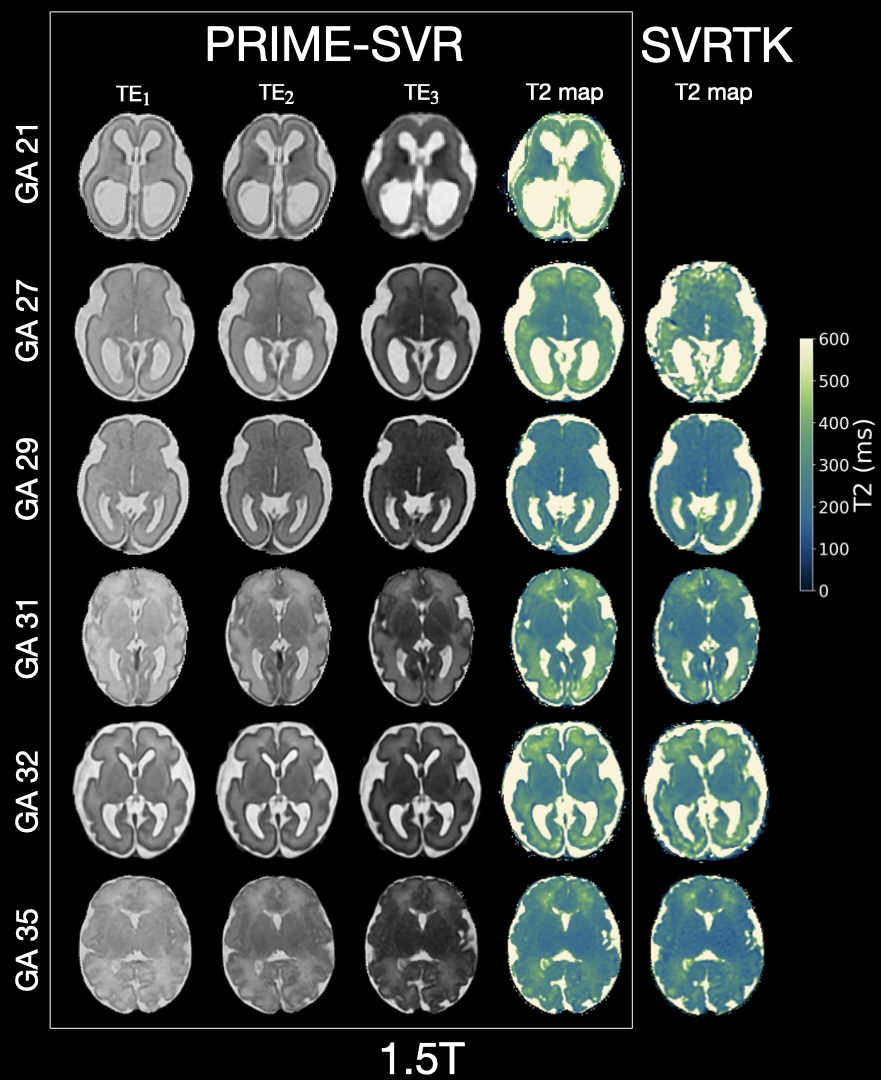}
     \caption{Overview of the HR reconstructions from PRIME-SVR (continued in Figure~\ref{fig:SRR_t22} across all subjects, along with the corresponding derived T2 maps (0.8 mm isotropic resolution). The last column shows the T2 maps derived from the SVRTK reconstructions.}
    \label{fig:SRR_t2}
    \end{figure}
    
\begin{figure}[bhpt]
    \centering
    \vspace{-1cm} 
     \includegraphics[width=0.7
    \textwidth]{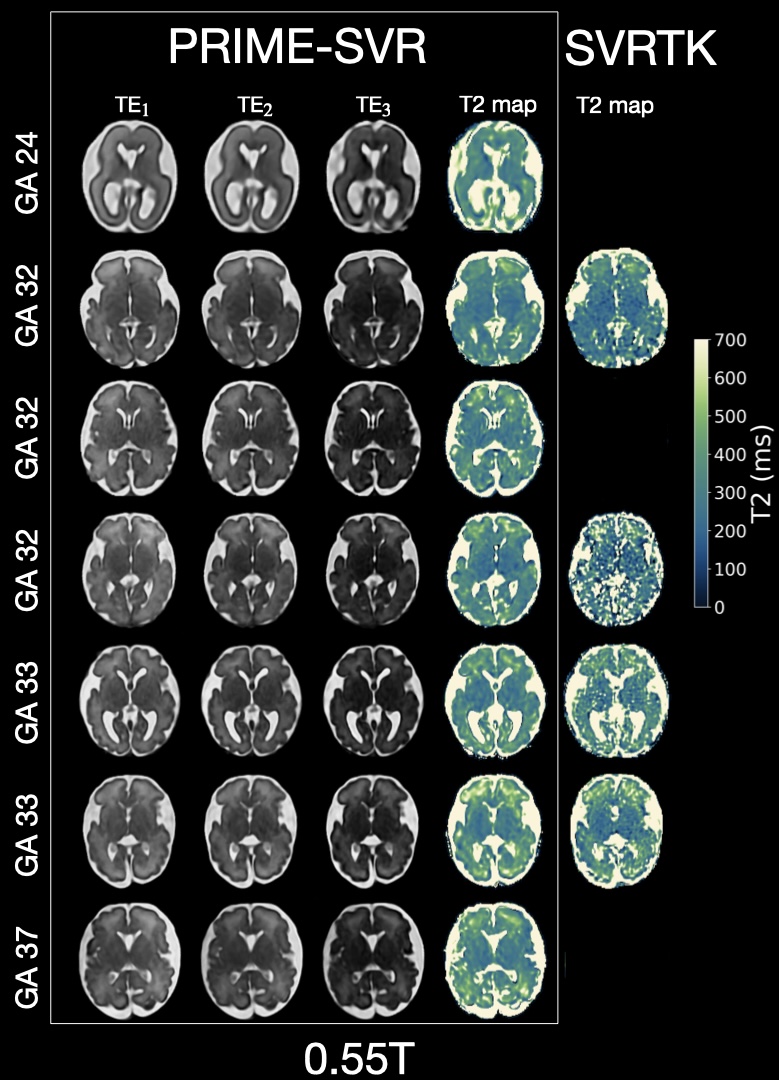}
\caption{Overview of the HR reconstructions from PRIME-SVR (continued from Fig.~\ref{fig:SRR_t2} across all subjects, along with the corresponding derived T2 maps (0.8 mm isotropic resolution). The last column shows the T2 maps derived from the SVRTK reconstructions.}
    \label{fig:SRR_t22}
\end{figure}
\subsection{Experiment 2: T2 mapping}
\label{exp:exp2}
\begin{table}[htbp]
\centering
\resizebox{\linewidth}{!}{
\begin{tabular}{llccc}
\toprule
Field strength & Method & WM (T2, ms) & DGM (T2, ms) & Residual fit (ms) \\
\midrule
\multirow{2}{*}{0.55~T} & PRIME-SVR* & 390 $\pm$ 79 & 269 $\pm$ 44 & 0.660 $\pm$ 1.32 \\
 & SVRTK & 379 $\pm$ 144 & 268 $\pm$ 100 & 3.11 $\pm$ 10.2 \\
\midrule
\multirow{2}{*}{1.5~T} & PRIME-SVR* & 336 $\pm$ 47 & 237 $\pm$ 23 & 4.01 $\pm$ 4.84 \\
 & SVRTK & 314 $\pm$ 50 & 222 $\pm$ 25 & 6.45 $\pm$ 16.9 \\
\bottomrule
\end{tabular}
}
\caption{Mean T2 (ms) $\pm$ mean intra-region standard deviation across subjects, for WM and DGM, and Residual fit reports the mean $\pm$ mean intra-subject standard deviation of the fit residual. *For fair comparison, only subjects with a successful SVRTK reconstruction are included for PRIME-SVR; over all subjects, PRIME-SVR values are 0.55\,T -- WM 384 $\pm$ 76, DGM 275 $\pm$ 46, and 1.5\,T -- WM 340 $\pm$ 48, DGM 241 $\pm$ 23.}
\label{tab:t2_values}
\end{table}

\begin{figure}[htbp]
    \centering
    \vspace{-1cm}
    \includegraphics[width=1\textwidth]{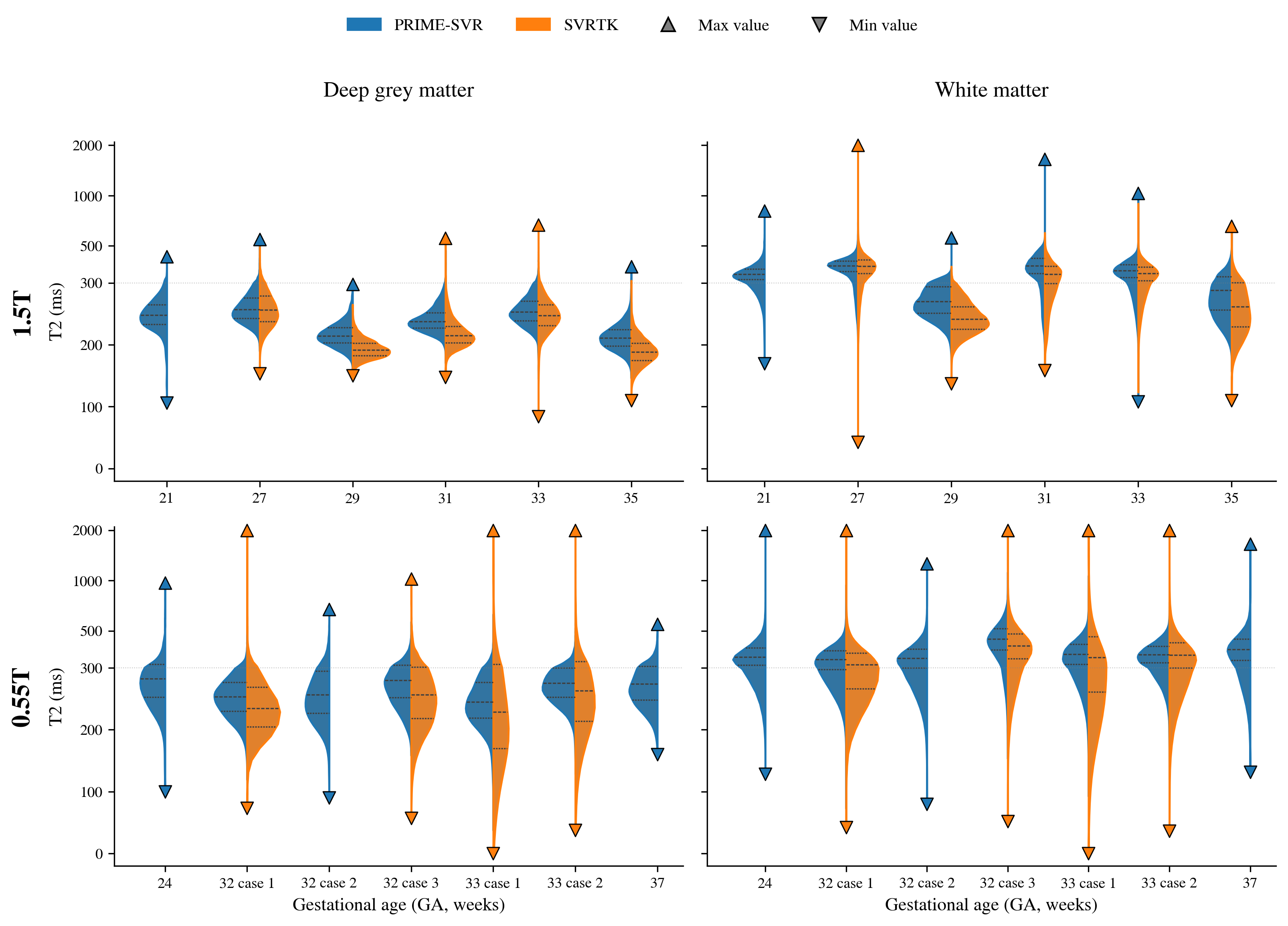}
    \caption{T2 distribution for each subject, obtained using PRIME SVR (blue) and SVRTK (orange). The first row corresponds to 1.5~T, the second row to low-field 0.55~T. The first column shows the deep gray matter, and the second column shows the white matter.}
    \label{fig:plott2}
\end{figure}

Figures~\ref{fig:SRR_t2} and~\ref{fig:SRR_t22} presents T2 maps obtained from PRIME-SVR and SVRTK reconstructions. Empty panels correspond to cases where T2 estimation could not be performed because at least one SVRTK reconstruction failed. 
T2 maps could not be derived from NeSVoR reconstructions because its reconstruction pipeline always rescales each volume's mean intensity to a fixed target value, with no option to disable this step, and this rescaling alters the relative signal decay across TEs.
PRIME-SVR enabled T2 estimation in four additional subjects (three at 0.55~T and one at 1.5~T) for whom SVRTK reconstruction failed. Table~\ref{tab:t2_values} summarizes the mean T2 values and intra-region variability measured in WM and DGM for the subset of subjects where both methods succeeded. Statistical comparisons were done in this subset. Mean T2 value in WM and DGM did not differ significantly between PRIME-SVR and SVRTK at either field strength whereas PRIME-SVR reduced the intra-region variability, particularly in WM (144\,ms to 79\,ms) and in DGM (100\,ms to 44\,ms) at 0.55~T. This is also visible qualitatively in Fig.~\ref{fig:SRR_t2}.~\ref{fig:SRR_t22} and Fig.~\ref{fig:plott2}, where PRIME-SVR generated more homogeneous T2 distributions across subjects, whereas SVRTK showed a wider distribution with more extreme values despite erosion to reduce CSF contamination. The fit residual was also not significantly different at either field strength, although PRIME-SVR showed a consistent direction of improvement, lower residual in all subjects at 0.55\,T and in 4 of 5 subjects at 1.5\,T. 

\begin{table}[tbp]
\centering
\caption{Comparison of the reconstruction quality when using less than 3 stacks per TE for PRIME-SVR.}
\vspace{-.2cm}
\resizebox{\linewidth}{!}{
\begin{tabular}{llccc}
\toprule
Field &\text{Input stacks} & \text{SSIM with HR} & \text{MAE** in T2} & \text{MAE in T2} \\
Strength& \text{number/TE} & \text{from 3 stacks/TE * $\uparrow$ } & \text{in WM $\downarrow$ } & \text{in DGM $\downarrow$ } \\
\midrule
\multirow{2}{*}{0.55T} &2 (6 subjects) & $0.884 \pm 0.053$ & $11.5 \pm 6.88$ & $0.544 \pm 0.258$ \\
&1  (5 subjects) & $0.859 \pm 0.069$ & $17.0 \pm 8.91$ & $0.844 \pm 0.279$ \\
\midrule
\multirow{2}{*}{1.5 T} &2 (5 subjects) &$0.920 \pm 0.042$ & $5.79 \pm 4.67$ & $0.291 \pm 0.263$ \\
&1 (4 subjects) & $0.883 \pm 0.043$ & $6.02 \pm 2.62$ & $0.373 \pm 0.120$ \\
\bottomrule
\end{tabular}}
\vspace{-.3cm}
\begin{flushleft}
\footnotesize{*Using full data, considered as reference reconstruction. **MAE: mean absolute error, WM: white matter, DGM: deep gray matter.}
\footnotesize{We discard subjects with $\bar{q} < 0.8$ for 2 stacks/TE, and $\bar{q} < 0.7$ for 1 stack/TE, since the reconstructions were not converging under this reduced setting.}
\end{flushleft}
\vspace{-.5cm}
\label{tab:red}
\end{table}

\subsection{Experiment 3: Robustness to limited input data}
\label{exp:exp3}
\begin{figure}[tbp]
    \centering
     \includegraphics[width=1\textwidth]{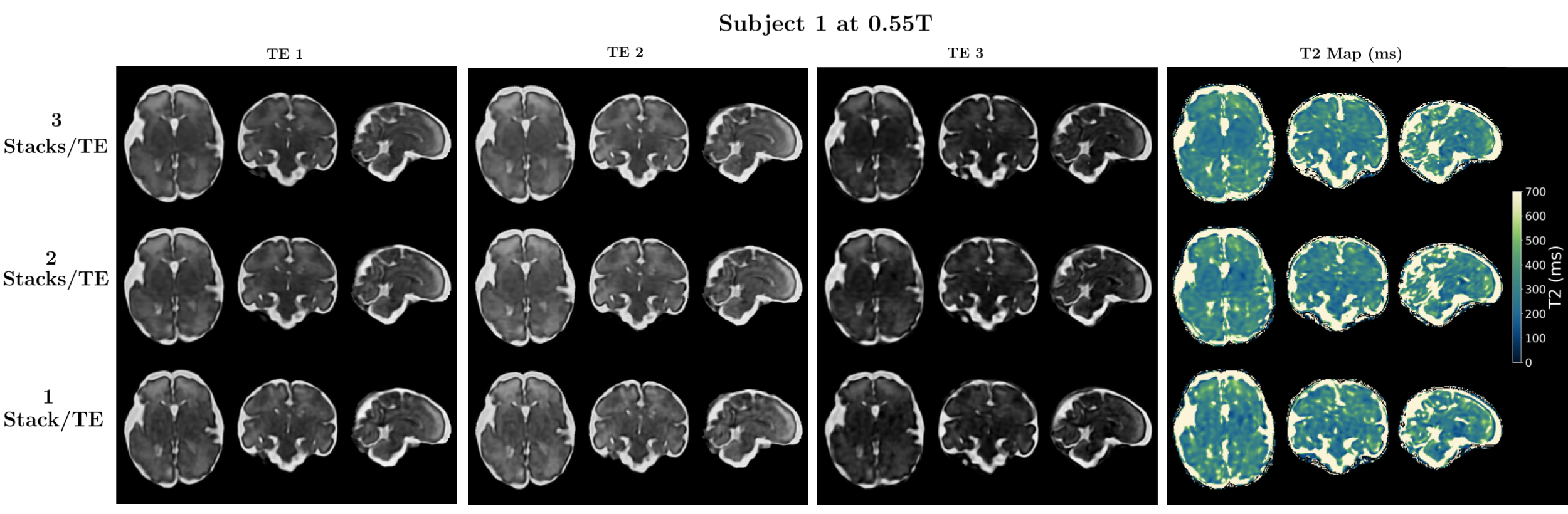}
    \vspace{0.1cm} 
     \includegraphics[width=1
    \textwidth]{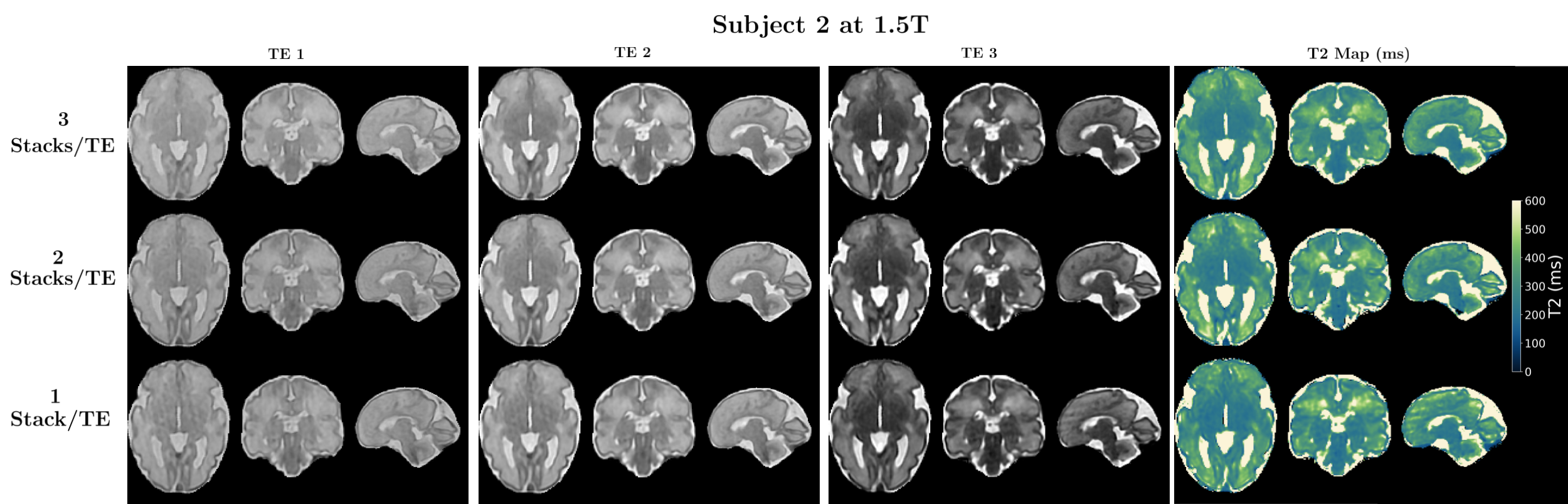}
\caption{HR reconstructions obtained with PRIME-SVR under reduced input data conditions (3, 2, and 1 stack per TE), shown across all TEs with corresponding T2 maps, for one representative subject at 0.55~T and 1.5~T.}
    \label{fig:red}
\end{figure}

 Table~\ref{tab:red} summarizes the reconstruction similarity metrics and the errors in T2 estimation within WM and DGM obtained using reduced numbers of input stacks per TE, with the HR volumes reconstructed from 3 stacks per TE (full data) serving as the reference. PRIME-SVR achieved higher similarity to the reference reconstruction (three stacks per TE) at 1.5~T than at 0.55~T. T2 errors were consistently lower at 1.5~T, particularly in WM. Across both field strengths, the mean absolute error in DGM remained below 1~ms for all reduced acquisition protocols. In contrast, WM errors increased more noticeably at 0.55~T, with a larger difference observed between the two- and one-stack-per-TE protocols. Representative reconstructions are shown in Fig.~\ref{fig:red}. 
Reconstructions obtained using 2 stacks/TE remain visually similar to the reference, whereas those reconstructed from a single stack per TE exhibit visible artifacts.

\subsection{Ablation Study}
\label{exp:ablation}
\subsubsection{Multi-TE Coupling}

Table~\ref{tab:ablation} summarizes the effect of a shared multi-echo network without Bloch regularization (Shared-SVR), independent networks per TE with Bloch regularization coupling (Reg-SVR) and the full model combining both (PRIME-SVR). PRIME-SVR did not differ significantly from Reg-SVR on AES or NMI, but significantly exceeded Shared-SVR on both. PRIME-SVR achieved the highest average Edge Dice across TE pairs, significantly outperforming both Reg-SVR and Shared-SVR.

\begin{table}[htbp]
\centering
\caption{Ablation of network sharing and Bloch regularization. AES is reported in $\times 10^{-3}$, averaged across TEs. NMI and edge Dice are reported as mean $\pm$ std across all TE pairs.}
\label{tab:ablation}
\begin{tabular}{lccc}
\toprule
 & \text{Reg-SVR} & \text{Shared-SVR} & \text{PRIME-SVR} \\
 & \scriptsize Single INR: $\times$, Reg.: $\checkmark$
 & \scriptsize Single INR: $\checkmark$, Reg.: $\times$
 & \scriptsize Single INR: $\checkmark$, Reg.: $\checkmark$ \\
 \midrule
AES $\uparrow$ &
$3.700 \pm 1.000$ &
$3.600 \pm 1.000$ &
$\mathbf{3.800 \pm 1.100}^{\dagger}$ \\
NMI $\uparrow$ 
&
$\mathbf{1.391 \pm 0.055}$ &
$1.359 \pm 0.042$ &
$1.388 \pm 0.061^{\dagger}$ \\
Edge Dice $\uparrow$ 
&
$0.551 \pm 0.110$ &
$0.458 \pm 0.132$ &
$\mathbf{0.619 \pm 0.121}^{*\dagger}$ \\
\bottomrule
\end{tabular}
\footnotesize{ $^{*}$/$^{\dagger}$ $p \leq 0.05$, exact two-sided Wilcoxon signed-rank test, paired by subject, comparing PRIME-SVR against Reg-SVR ($^{*}$) and Shared-SVR ($^{\dagger}$) separately, pooled across both field strengths ($n=13$; see Sec.~\ref{sec:stat}).}
\end{table}

\subsubsection{Adaptive weighting for Bloch regularization}

Table~\ref{tab:mixed_coeff_ablation} reports the effect of the regularization coefficient on T2 estimation in WM and DGM, reconstruction sharpness and cross-TE consistency. Increasing the fixed coefficient from 0.1 to 20 improved all reconstruction-quality metrics (NMI, Edge Dice, and AES), as the regularization enforces a stronger cross-TE signal consistency. However, this improvement came at the cost of a systematic upward bias in estimated T2 values, with WM T2 increasing from 353~ms to 377~ms and DGM T2 from 260~ms to 281~ms as the coefficient increased.  The adaptive weighting scheme mitigated this trade-off: it achieved T2 values close to those obtained with the lowest fixed coefficient (WM 357~ms, DGM 261~ms), while reaching reconstruction-quality metrics comparable to intermediate coefficients (NMI 1.39, Edge Dice 0.55, AES 3.76).

\begin{table}[t]
\centering
\caption{T$_2$ values and reconstruction quality metrics across Bloch regularization weights. AES is reported in units of $\times10^{-3}$.}
\label{tab:mixed_coeff_ablation}
\resizebox{\columnwidth}{!}{%
\begin{tabular}{lccccc}
\toprule
$\alpha$ & WM T$_2$ (ms) & DGM T$_2$ (ms) & NMI  $\uparrow$  & Edge Dice  $\uparrow$  & AES  $\uparrow$  \\
\midrule
0.1 & 354 $\pm$ 71.7 & 260 $\pm$ 44.2 & 1.38 $\pm$ 0.0360 & 0.514 $\pm$ 0.105 & 3.66 $\pm$ 1.00 \\
1   & 368 $\pm$ 73.2 & 269 $\pm$ 44.4 & 1.40 $\pm$ 0.0490 & 0.566 $\pm$ 0.110 & 3.63 $\pm$ 0.980 \\
10  & 374 $\pm$ 68.7 & 277 $\pm$ 40.1 & 1.44 $\pm$ 0.0440 & 0.614 $\pm$ 0.110 & 3.70 $\pm$ 0.990 \\
20  & 377 $\pm$ 72.0 & 282 $\pm$ 44.5 & 1.45 $\pm$ 0.0420 & 0.639 $\pm$ 0.104 & 3.84 $\pm$ 0.970 \\
\text{Adaptive} & \text{358 $\pm$ 70.6} & \text{261 $\pm$ 42.9} & \text{1.39 $\pm$ 0.0450} & \text{0.549 $\pm$ 0.113} & \text{3.76 $\pm$ 1.06} \\
\bottomrule
\end{tabular}%
}
\end{table}

\newpage
\section{Discussion}

\subsection{PRIME-SVR improves reconstruction robustness through self-supervised multi-TE reconstruction}
PRIME-SVR's performance stems from several intrinsic advantages. Jointly reconstructing across TEs lets the method leverage nine input stacks rather than three, increasing spatial coverage without additional acquisitions, an advantage especially valuable for late TEs, where SNR is low. PRIME-SVR also performed comparably at 0.55 T and 1.5 T and across TEs, underscoring its self-supervised design's ability to generalize across acquisition settings (Sec. \ref{exp:exp1}). Both NMI and Edge Dice captured the temporal continuity of reconstructions across TEs and increased as we strengthened the coefficient.
This multi-TE coupling, however, can bias the derived T2 maps: overly strong Bloch regularization trades quantitative accuracy for reconstruction consistency, so the two must be balanced. Adaptive weighting appears to recover this balance (Sec. \ref{exp:ablation}).
These results are notable given two major domain shifts, magnetic field strength and TE, for which neither NeSVoR nor SVRTK was originally designed or trained. Both methods nonetheless performed reasonably well overall, though both struggle at late TEs, which hamper their ability to be used for T2 mapping. 

\subsection{Accurate reconstruction is essential for reliable quantitative T2 mapping}

Accurate reconstruction, including preservation of the underlying signal intensities, is essential for reliable quantitative T2 mapping. To date, and to our knowledge, SVRTK has been the only SVR framework applied to multi-echo fetal MRI reconstruction for T2 mapping. The applicability of NeSVoR to quantitative mapping is limited by the absence of an option to preserve the original intensity range. Here, we demonstrate, for the first time, T2 mapping from INR-based reconstructions, achieving an isotropic resolution of 0.8 mm (Sec.\ref{exp:exp2}). The T2 values obtained with PRIME-SVR and SVRTK are nevertheless higher than those previously reported using dictionary fitting based on signals simulated with the EPG formalism \citep{bhattacharya_vivo_2024}. This discrepancy is expected from the simple monoexponential model used in our study, which does not account for slice-profile effects and consequently leads to T2 overestimation, as previously demonstrated in phantom and adult experiments \citep{lajous_fetal_2022,roulet_t2_2025}. T2 values were longer at 0.55 T than at 1.5 T, consistent with the field-dependence of T2 relaxation. Finally, we did not observe the expected decrease in T2 with GA. However, in this study, T2 values were reported over relatively large regions, such as WM and DGM, whereas GA-related T2 changes has been reported within more spatially specific WM regions \citep{ge_highfidelity_2026}. In addition, the limited number of subjects available at each GA in our cohort precludes a robust assessment of age-related trajectories. In future work, a larger cohort combined with a more spatially resolved regional analysis will therefore be required to study in depth developmental T2 changes in normal and abnormal brains.

\subsection{Towards faster quantitative fetal MRI}

PRIME-SVR reduces the time required for quantitative fetal MRI at both the reconstruction and acquisition stages. In our experimental setting, SVRTK was particularly fast (2min) due to the availability of 32 CPU cores, although reported runtimes in other settings are comparable to, or slightly slower than, INR based methods \citep{dannecker2026fast}. By jointly reconstructing all TEs, PRIME-SVR requires only 5 min for total reconstruction time, compared with 21 min for NeSVoR (7 min per TE) (Sec. \ref{exp:exp1}). Joint reconstruction also reduces acquisition time. By exploiting information across TEs, high-resolution T2 maps can be obtained from only two stacks per TE, or even one orthogonal stack per TE when image quality is sufficient (Sec. \ref{exp:exp3}). This reduces acquisition time from approximately 15 to 5 min, addressing a major barrier to the clinical adoption of quantitative volumetric fetal MRI \citep{bhattacharya_vivo_2024}.

\subsection{Limitations and future work}

Several limitations deserve mention. First, the dataset remains modest at 39 acquisitions (13 subjects with 3 TEs each), although it is currently the largest existing in vivo multi-echo fetal dataset. The scarcity of such datasets may partly reflect the challenges of reconstructing late-TE fetal data with existing SVR methods. Second, the regularization relies on a simplified Bloch signal model. More sophisticated models, such as EPG \citep{weigel_extended_2015}, could be used in different ways. First, EPG-simulated dictionary fitting could correct the T2 overestimation observed here, as suggested in \cite{bhattacharya_vivo_2024}. This would, however, require EPG simulations incorporating the specific sequence parameters of our low-field acquisition. Second, EPG could be incorporated directly into the continuous function learned by the INR to regress T2 values directly. Third, on the methodological side, Gaussian splatting could become a strong alternative to INRs: by computing the PSF convolution analytically, it avoids the discretization errors introduced by Monte Carlo sampling~\citep{dannecker2026fast}.
Finally, our framework is not yet fully automated. In pre-processing steps, we generated brain masks for the second and third TEs manually when existing methods failed; automating this step would be necessary for analyzing larger cohorts and to support clinical deployment.

\section{Conclusion}

We presented PRIME-SVR, the first implicit neural representation framework for joint slice-to-volume reconstruction of multi-echo fetal brain MRI, enabling self-supervised, protocol-independent quantitative T2 mapping. By coupling a continuous spatio-temporal signal representation with a Bloch equation-derived cross-TE regularization, PRIME-SVR enforces physically consistent T2 decay across TEs and remains robust to the motion and degradation typical of fetal acquisitions. Evaluated on the largest multi-centric, multi-vendor, multi-field-strength cohort of multi-TE fetal MRI to date, PRIME-SVR outperformed state-of-the-art SVR methods developed for single-TE clinical acquisitions, in sharpness, anatomical accuracy, and cross-TE consistency. Notably, it extended reconstruction to late TEs and low field strength previously inaccessible to SVR, yielding the first sub-millimeter isotropic T2 maps at 0.55 T and the first T2 maps derived from an INR-based reconstruction. Beyond reconstruction quality, PRIME-SVR maintains accurate T2 estimation from significantly fewer acquired stacks, allowing meaningful reductions in acquisition time with minimal loss in accuracy. This directly addresses one of the main practical barriers to the clinical translation of quantitative fetal MRI, bringing T2 mapping closer to routine, time-constrained fetal protocols.
\looseness=-1
We hope this work encourages the community to acquire and share more multi-TE fetal datasets. Although our focus here is on early brain development, PRIME-SVR's self-supervised, physics-informed design extends beyond the fetal brain: it could apply to other applications combining super-resolution and quantitative mapping, such as musculoskeletal T2 mapping, or, with application-specific non-rigid motion modeling, to liver or cardiac MRI.

\section{Acknowledgments}
This research was funded by the Swiss National Science Foundation (205320-215641; 220442), by the ERC (Deep4MI - 884622), and
by the ERA-NET NEURON (MULTI-FACT - 8810003808, SNSF 31NE30 203977). 
The project leading to this publication has received funding from the Excellence Initiative of Aix-Marseille Université - A*Midex, a French “Investissements d’Avenir programme” AMX-21-IET-017. Furthermore, this project was supported by the French Agence Nationale de la Recherche (grants ANR-19-CE45-0014, ANR-21-NEU2-0005) and La Fondation de France (n°00147743).
This work was supported by the Wellcome Trust through a Sir Henry Wellcome Fellowship [201374/Z/16/Z], the UKRI Future Leaders Fellowship [MR/T018119/1], the DFG Heisenberg Programme [502024488], the ERC Starting Grant EARTHWORM [101165242], the NIHR Advanced Fellowship [NIHR3016640], MRC grants [MR/W019469/1] and [MR/X010007/1], and the Wellcome/EPSRC Centre [WT203148/Z/16/Z]. We acknowledge access to the facilities and expertise of the CIBM Center for Biomedical Imaging, a Swiss research center of excellence founded and supported by CHUV, UNIL, EPFL, UNIGE and HUG.




\newpage

\bibliographystyle{plainnat}

\bibliography{references_modif}

\newpage
\appendix
\section{PSF definition}
\label{sec:annexe1}
$M_{j,k}(\mathbf{x})$ is the anisotropic Gaussian point spread function (PSF) centered at the spatial location corresponding to pixel $k$ in slice $j$, denoted $\mathbf{p}_{j,k}$. It is modeled as an anisotropic Gaussian:
For clarity of exposition, the detailed formulation of the PSF, 

\begin{equation*}
M_{j,k}(\mathbf{x})
=
\frac{1}{(2\pi)^{3/2} \det(\boldsymbol{\Sigma})^{1/2}}
\exp\!\left(
-\frac{1}{2}
\left(
T_j^{-1}(\mathbf{x}) - \mathbf{p}_{j,k}
\right)^\top
\boldsymbol{\Sigma}^{-1}
\left(
T_j^{-1}(\mathbf{x}) - \mathbf{p}_{j,k}
\right)
\right),
\end{equation*}
where $T_j$ is the unknown rigid transformation (rotation and translation) mapping slice $j$ into the HR 3D space. The covariance matrix $\boldsymbol{\Sigma}$ is defined following~\citep{rousseau_registration-based_2006} as

\begin{equation*}
\boldsymbol{\Sigma}
=
\mathrm{diag}
\left(
\left(\frac{1.2\, r_1}{2.355}\right)^2,
\left(\frac{1.2\, r_2}{2.355}\right)^2,
\left(\frac{r_3}{2.355}\right)^2
\right),
\end{equation*}
where $r_1$ and $r_2$ denote the in-plane pixel spacings and $r_3$ denotes the slice thickness.

\section{Fisher Information}
\label{sec:annexe2}
Assuming a mono-exponential signal $S(t;\mathrm{T2})=M_0 e^{-t/\mathrm{T2}}$ corrupted by Gaussian noise $x=S+\epsilon$, $\epsilon\sim\mathcal N(0,\sigma^2)$, the likelihood is
\[
p(x;\mathrm{T2})\propto \exp\left(-\frac{(x-S)^2}{2\sigma^2}\right).
\]
The Fisher information is defined as
\[
\mathcal I(\mathrm{T2})=\mathbb{E}\left[\left(\frac{\partial}{\partial \mathrm{T2}}\log p(x;\mathrm{T2})\right)^2\right].
\]
Differentiating the log-likelihood yields
\[
\frac{\partial}{\partial \mathrm{T2}}\log p(x;\mathrm{T2})
=\frac{1}{\sigma^2}(x-S)\frac{\partial S}{\partial \mathrm{T2}}.
\]
Taking the expectation and using $\mathbb{E}[(x-S)^2]=\sigma^2$ gives
\[
\mathcal I(\mathrm{T2})=\frac{1}{\sigma^2}\left(\frac{\partial S}{\partial \mathrm{T2}}\right)^2.
\]
With $\frac{\partial S}{\partial \mathrm{T2}}=M_0 \frac{t}{\mathrm{T2}^2} e^{-t/\mathrm{T2}}$, we obtain
\[
\mathcal I(t)=\frac{M_0^2}{\sigma^2}\frac{t^2}{\mathrm{T2}^4} e^{-2t/\mathrm{T2}}.
\]

To determine the echo time by maximizing the Fisher information for a single measurement, we maximize the term
\begin{equation*}
f(t) = t^2 e^{-2t/\mathrm{T2}}.
\end{equation*}
Differentiating gives
\begin{equation*}
f'(t) = 2t e^{-2t/\mathrm{T2}} - \frac{2}{\mathrm{T2}} t^2 e^{-2t/\mathrm{T2}}
= 2t e^{-2t/\mathrm{T2}} \left(1 - \frac{t}{\mathrm{T2}}\right).
\end{equation*}
Setting $f'(t)=0$ yields $t=0$ or $
t = \mathrm{T2}.$
Since $t=0$ corresponds to zero Fisher information, the maximum is achieved at
\begin{equation*}
t = \mathrm{T2}.
\end{equation*}
Thus, the Fisher information is maximized when the TE is sampled close to the expected T2 value.

\end{document}